\documentclass[a4paper,fleqn,usenatbib]{mnras}
\usepackage[T1]{fontenc}
\usepackage{ae,aecompl}
\usepackage{subfig}

\usepackage{graphicx}	% Including figure files
\usepackage{amsmath}	% Advanced maths commands
\usepackage{amssymb}	% Extra maths symbols

\title[Major Mergers II: Profile Changes]{Major mergers between dark matter haloes -- II. Profile and Concentration Changes}

\author[N.E. Drakos et al.]{
	Nicole E. Drakos$^{1,2} $\thanks{E-mail: ndrakos@uwaterloo.ca},
	James E. Taylor$^{1,2}$,
	Anael Berrouet$^{2}$,
	\newauthor
	Aaron S. G. Robotham$^{3}$,
	Chris Power$^{3}$,
	\\
	$^{1}$Waterloo Centre for Astrophysics, University of Waterloo, 200 University Avenue West, Waterloo, ON, N2L\,3G1, Canada \\	
	$^{2}$Department of Physics and Astronomy, University of Waterloo, 200 University Avenue West, Waterloo, ON, N2L\,3G1, Canada \\
	$^{3}$ ICRAR, University of Western Australia, 35 Stirling Highway, Crawley, Western Australia 6009, Australia \\
}

\date{Accepted XXX. Received YYY; in original form ZZZ}

\pubyear{2019}

\begin{document}
	\label{firstpage}
	\pagerange{\pageref{firstpage}--\pageref{lastpage}}
	\maketitle
	
\begin{abstract}
Several lines of evidence suggest that as dark matter haloes grow their scale radius increases, and that the density in their central region drops. Major mergers seem an obvious mechanism to explain both these phenomena, and the resulting patterns in the concentration--mass--redshift relation. To test this possibility, we have simulated equal-mass mergers between haloes with a variety of cosmological density profiles, placed on various different orbits.  The remnants typically have higher densities than the initial conditions, but differ only slightly from self-similar scaling predictions. They are reasonably well fit by Einasto profiles, but have parameters distinct from those of the initial conditions. The net internal energy available to the merger remnant, relative to the internal energy of the initial conditions, $\kappa$, has the greatest influence on the properties of the final mass distribution. As expected, energetic encounters produce more extended remnants while mergers of strongly bound systems produce compact remnants. Surprisingly, however, the scale radius of the density profile shows the opposite trend, {\it increasing} in the remnants of low-energy encounters relative to energetic ones. Also even in the most energetic encounters, the density within the scale radius decreases only slightly (by 10--20\%), while for very low-energy systems it increases significantly after the merger. We conclude that while major mergers can produce remnants that are more diffuse at large radii, they are relatively ineffective at changing the central densities of haloes, and seem unlikely to explain the mean trends in the concentration--mass--redshift relation.
\end{abstract}

	
%%%%%%%%%%%%%%%%%%%%%%%%%%%%%%%%%%%%%%%%%%%%%%%%%%
\section{Introduction}

Dark matter haloes play a central role in our current understanding of cosmological structure formation, being the site of all visible galaxy formation. While observational tests including galaxy kinematics \citep[e.g.][]{ouellette2017}, satellite kinematics \citep[e.g.][]{guo2012}, and weak and strong gravitational lensing \citep[e.g.][]{umetsu2016} place important constraints on halo properties, most of our detailed knowledge of halo structure comes from numerical simulations. A remarkable property of dark matter haloes discovered in cosmological simulations is the existence of a nearly universal density profile, regardless of mass or cosmological model \citep{navarro1996,navarro1997}, commonly approximated by the Navarro--Frenk--White (NFW) form,
\begin{equation}
\rho(r) = \dfrac{\rho_0 r_{\rm s}^3}{r(r+r_{\rm s})^2} \,\,\, ,
\end{equation}
where $\rho_0$ is a characteristic density and $r_{\rm s}$ is the scale radius, describing the point where the logarithmic slope is ${\rm d} \ln \rho/ {\rm d} \ln r = -2$.  Though NFW profiles are still commonly used in the literature, dark matter halo density profiles are slightly better described by Einasto profiles 
%\citep[e.g.][]{navarro2004, merritt2006,gao2008, springel2008, navarro2010, ishiyama2013,dutton2014,klypin2016}
\citep[e.g.][]{navarro2004, gao2008,klypin2016}, which can be expressed as:
\begin{equation}
\rho(r) = \rho_{-2} \exp \left(-\dfrac{2}{\alpha_{\rm E}}\left[\left(\dfrac{r}{r_{-2}}\right)^{\alpha_{\rm E}} - 1 \right]\right) \,\,\, , 
\end{equation}
where $\alpha_{\rm E}$ is the Einasto shape parameter and $r_{-2}$ is the radius where the logarithmic slope is $-2$. 

The mean density of dark matter haloes within their outer, virial boundary scales with the mean or critical background density at the epoch at which they are observed, but is the same for all haloes at that epoch, independent of mass or growth history. On the other hand, the central densities of haloes at any one epoch can vary considerably, depending on their concentration. The concentration parameter $c$ was originally defined in terms of the NFW density profile as $c = {r_{\rm vir}}/{r_{\rm s}}$, where $r_{\rm vir}$ is the virial radius. This definition can be extended to the Einasto profile by taking $r_{\rm s} = r_{-2}$; this definition does not capture the effects of $\alpha_{\rm E}$ on the central density of the halo \citep{klypin2016}, however. An alternative, profile independent, definition of concentration is based on the ratio of the maximum circular velocity to the virial velocity $v_{\rm peak}/v_{\rm vir}$ \citep{prada2012, klypin2016}, which is monotonically related to the original definition of $c$.

For an individual halo, the evolution of the concentration parameter is linked to the halo's merger history \cite[e.g.][]{navarro1997, bullock2001, vandenbosch2002, zhao2009, wong2012, klypin2016}. 
%{navarro1997, jing2000, bullock2001, wechsler2002, vandenbosch2002, zhao2003a, zhao2003b, tasitsiomi2004, neto2007, duffy2008, gao2008, maccio2008, zhao2009, prada2012, wong2012, ishiyama2013, dutton2014, diemer2015, klypin2016}. 
These previous studies have established that haloes undergo two main phases of mass accretion. In the first, rapid phase, the concentration parameter remains roughly constant, with a value of $c \approx 3$. In the second, slow phase, the concentration parameter grows as the virial radius increases while the scale radius remains fixed. The increase in mass and concentration during this second phase is mainly due to the decreasing reference density, and is therefore sometimes referred to as pseudo-evolution \citep[e.g.][]{diemer2013}. Averaged over many systems, these patterns give rise to the mean concentration--mass--redshift relation, in which concentration generally decreases with increasing mass \citep[e.g.][]{navarro1996,navarro1997, jing2000, bullock2001}, except at the largest masses, where velocity-based definitions can increase again \citep[e.g.][]{klypin2016}.

An important implication of previous measurements of the concentration--mass--redshift relation is that the inner scale radius must increase as haloes grow. In cosmological simulations, the median concentration of haloes of a given mass evolves with redshift as $c \sim c_0 (1+z)^{-1}$, as first demonstrated by  \cite{bullock2001}, or possibly as $c \sim c_0\rho_{\rm c}^{-1/3}$,  (where $\rho_{\rm c}$ is the critical density of the universe), as pointed out by \cite{pilipenko2017}. This would correspond to the scaling of the virial radius for an object whose mass did not increase with redshift, that is for an isolated system surrounded by a void. Haloes never exist in complete isolation, however; as the virial radius increases, it will enclose more matter, increasing both the total mass and the virial radius further. Thus, the net growth will go as $r_{\rm vir} \propto (M_{\rm vir} / \rho_{\rm c})^{1/3}$. For concentration to scale as $c \propto \rho_{\rm c}^{-1/3}$, the scale radius must therefore increase as $M_{\rm vir}^{1/3}$ as well. A similar conclusion was reached by \cite{zhao2003a}, who showed directly that $r_{\rm s}$ increases during the period of rapid accretion. Provided $r_{\rm s}$ increases as $M_{\rm vir}^{1/3}$, the density at or within one scale radius will remain constant, while the density within a {\it fixed physical} radius increases. 

This prediction seems puzzling given several other pieces of evidence that central density must decrease as haloes grow. The first was discovered by  \cite{nusser1999}, who showed that in the absence of any rearrangement of the pre-existing material, accretion onto the outside of a halo would produce a structure with a central density much higher than that of haloes in cosmological simulations. A second piece of evidence comes from simulations of the first haloes by \cite{ishiyama2014}. Evolving these down to a final redshift of $z=32$, he found central densities that were once again much higher at fixed physical radius than expected from extrapolations of the low-redshift concentration--mass--redshift relations; if these densities are conserved, they would increase estimates of the boost factor by up to two orders of magnitude \citep{okoli2018}. From both these studies, the implication is that there must be some mechanism that rearranges the central parts of haloes, causing the mass distribution to expand, and decreasing the central density. Given the work of \cite{nusser1999}, this mechanism must be distinct from slow accretion and associated with periods of rapid growth; thus major mergers seem an obvious candidate.

Isolated major mergers between otherwise undisturbed systems are rare in a cosmological context, and thus the effects of major mergers are best studied through controlled simulations of isolated systems \citep[e.g.][]{boylankolchin2004, aceves2006, kazantzidis2006,mcmillan2007, zemp2008, vass2009, ogiya2016,drakos2018}.
%{white1978,farouki1983, villumsen1983, barnes1992,fulton2001,moore2004,boylankolchin2004, aceves2006, kazantzidis2006,valluri2007,zemp2008, vass2009, ogiya2016,drakos2018}.
Most authors find that haloes are robust to major mergers; in particular, it appears that the slope of the steepest density profile is preserved \citep{boylankolchin2004, aceves2006, kazantzidis2006,zemp2008,vass2009}, and there is also some suggestion that concentration does not change \citep{kazantzidis2006}. However, the simulations of \cite{moore2004} suggest that concentration can decrease in major mergers, provided the initial encounter has sufficient angular momentum. Furthermore, the simulations of \cite{ogiya2016} suggest that if the inner slopes of the dark matter haloes are particularly steep (as in the case of primordial haloes), major mergers can also cause a decrease in the inner slope (again with some orbital dependence).

In a recent study \citep[][hereafter Paper I]{drakos2018}, we performed a large number of isolated major merger simulations, covering a wide range of initial halo models and orbital parameters to study how the spin, size, and 3D shape of merger remnants depend on the initial conditions (ICs) and the orbital parameters describing the encounter. We showed that the spin and shape of merger remnants depend mainly on the angular momentum and energy of the merger. In this paper, we will consider how the characteristic radii, density profile, and concentration parameter depend on orbital properties and ICs. Our main goal is to determine whether major mergers are a viable mechanism for increasing scale radius and/or decreasing central density, and if so, under which conditions. 

The outline of the paper is as follows: in Section~\ref{sec:Sims}, we briefly review our method for generating ICs, and the merger simulations from Paper I. In Section~\ref{sec:SelfSim}, we examine the overall behaviour of the density profile and various characteristic radii. In Section~\ref{sec:ProfFit}, we show the results of fitting analytic density profiles to the remnants, and in Section~\ref{sec:Conc}, we consider the implications for the concentration parameter. We summarize and discuss our results in Section~\ref{sec:Discuss}.

%%%%%%%%%%%%%%%%%%%%%%%%%%%%%%%%%%%%%%%%%%%%%%%%%%%%%%%%%%%%%%%%%%%
\section{Simulations} \label{sec:Sims}

In this section we briefly outline the merger simulations. For a more detailed explanation of the simulations, see Paper I.

\subsection{Initial profile models}

The halo ICs were created using the public code \textsc{icicle} \citep{drakos2017}. We considered six different halo models. Two were NFW models truncated using an exponential cut-off (NFWX), and two were NFW models truncated by iteratively removing unbound particles outside a specified radius (NFWT). The last two models were Einasto profiles (Ein), one with a low shape parameter $\alpha_{\rm E}=0.15$, and one with a high shape parameter $\alpha_{\rm E}=0.3$. These values of $\alpha_{\rm E}$ span the range found in cosmological simulations by \cite{gao2008}. The Einasto profiles were not truncated, as the total mass converges to a finite value with increasing radius. The properties of the profiles are summarized in Table~\ref{tab:ICs}. The simulation units were chosen so that the gravitational constant, $G$, the peak circular velocity, $v_{\rm peak}$, and the radius at which the circular velocity peaks, $r_{\rm peak}$, are all unity. Setting $G=M_{\rm peak}=r_{\rm peak}=1$ produces a time unit $t_{\rm unit} = \sqrt{r_{\rm peak}^3/GM_{\rm peak}}$, a density unit $\rho_{\rm unit}=M_{\rm peak}/r_{\rm peak}^3$, and an energy unit $E_{\rm unit} = GM_{\rm peak}^2/r_{\rm peak}$. 

\begin{table*} 
	\caption{\label{tab:ICs}Summary of IC properties. The columns list (1) the name of the IC, (2) the number of particles $N$, (3) the parameters used to construct the IC, and (4) the total internal energy of the IC, $E_0$.}
	\begin{tabular}{c c c c c}
		\hline
		Initial condition name & $N$ &Parameters & $E_0/E_{\rm unit}$\\ 
		\hline
		EinLow & $5 \times 10^5$& $\alpha_{\rm E}=0.15$ &  -2.2\\
		EinHigh& $5 \times 10^5$ & $\alpha_{\rm E}=0.3$ &  -1.2\\
		NFWT10& $\sim 3.2 \times 10^5$ & $r_{cut} = 10$  & -1.0\\	
		NFWT15& $ \sim 3.5  \times 10^5$ & $r_{cut} =15$  &  -1.3\\	
		NFWXSlow& $5 \times 10^5$ & $r_{\rm vir}=10$, $r_{d} = 2 \, r_{\rm s}$&  -1.6\\	
		NFWXFast& $5 \times 10^5$ &  $r_{\rm vir}=10$, $r_{d} = 0.2 \, r_{\rm s}$&  -1.5\\			
		\hline
	\end{tabular}	
\end{table*}

\subsection{Orbits}

For each of the six density profiles, we simulated encounters between two identical haloes with that profile, the first initially at rest and the second on a specific initial orbit. We considered 29 different orbits: 15 with a purely radial (R), and 14 with a purely tangential (T) initial velocity in the frame of the first halo. The haloes had an initial separation of $r_{\rm sep}=2$, $5$, or $10\,r_{\rm peak}$ and an initial velocity of $v_0 =  0.1,\, 0.2,\, 0.6,\, 0.8 \, $, or $1.2 \, v_{\rm esc}$, where $v_{\rm esc}$ is the escape speed of a point mass located at $r_{\rm sep}$. For $v_0=1.2\, v_{\rm esc}$, tangential orbits do not produce a bound remnant, so we did not simulate tangential orbits with this highest velocity. As described in Paper I, the merger simulations were run in \textsc{gadget-2} \citep{gadget2} using a softening length of $\epsilon= 0.02\, r_{\rm peak}$. The centre of the remnant halo was found by calculating the centre of mass within increasingly smaller spheres. 

In Paper I, we found that the 3D shape of the final merger remnant depends on the energy and angular momentum of the initial orbit. The dependence is simplest when expressed in terms of dimensionless energy and spin parameters $\kappa$ and $\lambda$:  
\begin{equation}
	\begin{aligned}
		\kappa &= \dfrac{E_0'}{E_0} \left(\dfrac{M}{M'}  \right)^{5/3} \\
		\lambda &= \dfrac{\sqrt{|E_0'|}}{GM'^{5/2}}|\textbf{J}'| \,\,\, .
	\end{aligned}
\end{equation}
Here $E_0'$ and $E_0$ are the internal energies of the remnant and of the initial halo, while $M'$ and $M$ are their respective masses. The definition of the energy parameter includes the factor $(M/M')^{5/3}$, since the energy scales as $E_0 \propto M^2/ r \propto M^{5/3}$ for `self-similar' growth, that is growth that conserves the mean density and the form of the density profile, as explained in Paper I. Given this definition, $\kappa=1$ corresponds to a final remnant that is self-similar to the ICs, $\kappa<1$ indicates that the remnant is less bound than the progenitor, and $\kappa>1$ indicates it is more bound. The dimensionless spin parameter follows the definition of \cite{peebles1969}, as discussed in Paper I. 

\begin{figure}
	\includegraphics[width = \columnwidth,trim={3cm 4.5cm 3.5cm 0},clip]{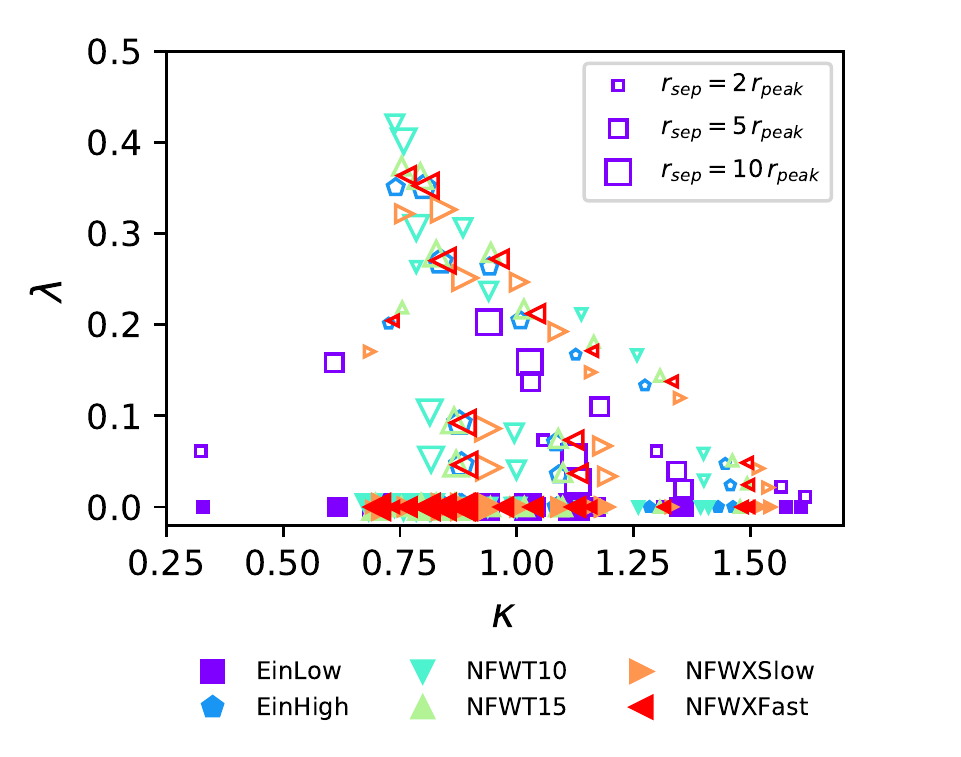}.
	\caption{The range of orbital parameters used in the simulations. The orbital parameters are the energy parameter $\kappa$ (which expresses the change in internal energy from the ICs to the final remnant, relative to the self-similar value for a binary merger), and the spin parameter, $\lambda$. The open points indicate tangential initial velocities, and filled points denote radial initial velocities. The size of the symbols corresponds to the initial radial separation, $r_{\rm sep}$.
	}	
	\label{fig:OrbitalParameters2}
\end{figure}

The range of $\kappa$ and $\lambda$ used is shown in Fig.~\ref{fig:OrbitalParameters2}. In this space, there is a natural restriction on the spin parameter for both small and large values of $\kappa$. For small values of $\kappa$, orbits with larger $\lambda$ values become unbound, while large $\kappa$ simulations correspond to tightly bound orbits with small radial separations and velocities, so the range of $\lambda$ is also limited.

Conveniently, both $\kappa$ and $\lambda$ can be calculated directly from the ICs using conservation laws. Specifically, the angular momentum $\mathbf{J}' $ and internal energy $E_0'$ of the remnant can be calculated given the radial separation $r_{\rm sep}$, the initial velocity $v_0$, the orbital energy $E_{\rm orb}$, the internal energy of the initial halo $E_0$ and the mass of the initial halo $M$:
\begin{equation} \label{eq:JandE}
\begin{aligned}
\mathbf{J}' &= M \mathbf{r_{\rm sep}} \times \mathbf{v_0}\,\,\, , \\
E_0' &= E_{\rm orb}+2 E_0 - \dfrac{1}{4} M v_0^2 \,\,\,.
\end{aligned}
\end{equation}
In Paper I, it was found that $E_0'$ and $J'$ calculated in this manner agreed to a direct calculation to within 2\% and 0.5\%, respectively.

%%%%%%FIGURE 1%%%%%%

 %%%%%%%%%%%%%%%%%%%%%%%%%%%%%%%%%%%%%%%%%%%%%%%%%%%%%%%%%%%%%%%%%%%
\section{Net Change in the Mass Distribution} \label{sec:SelfSim}

In this section, we explore how the density profile changes going from the IC to the final remnant. As in Paper I, the properties of the final remnant are measured at time $300\, t_{\rm unit}$, by which time it has fully relaxed.

\subsection{General results}

\begin{figure*}
	\includegraphics[trim={0 1cm 0 1cm},clip]{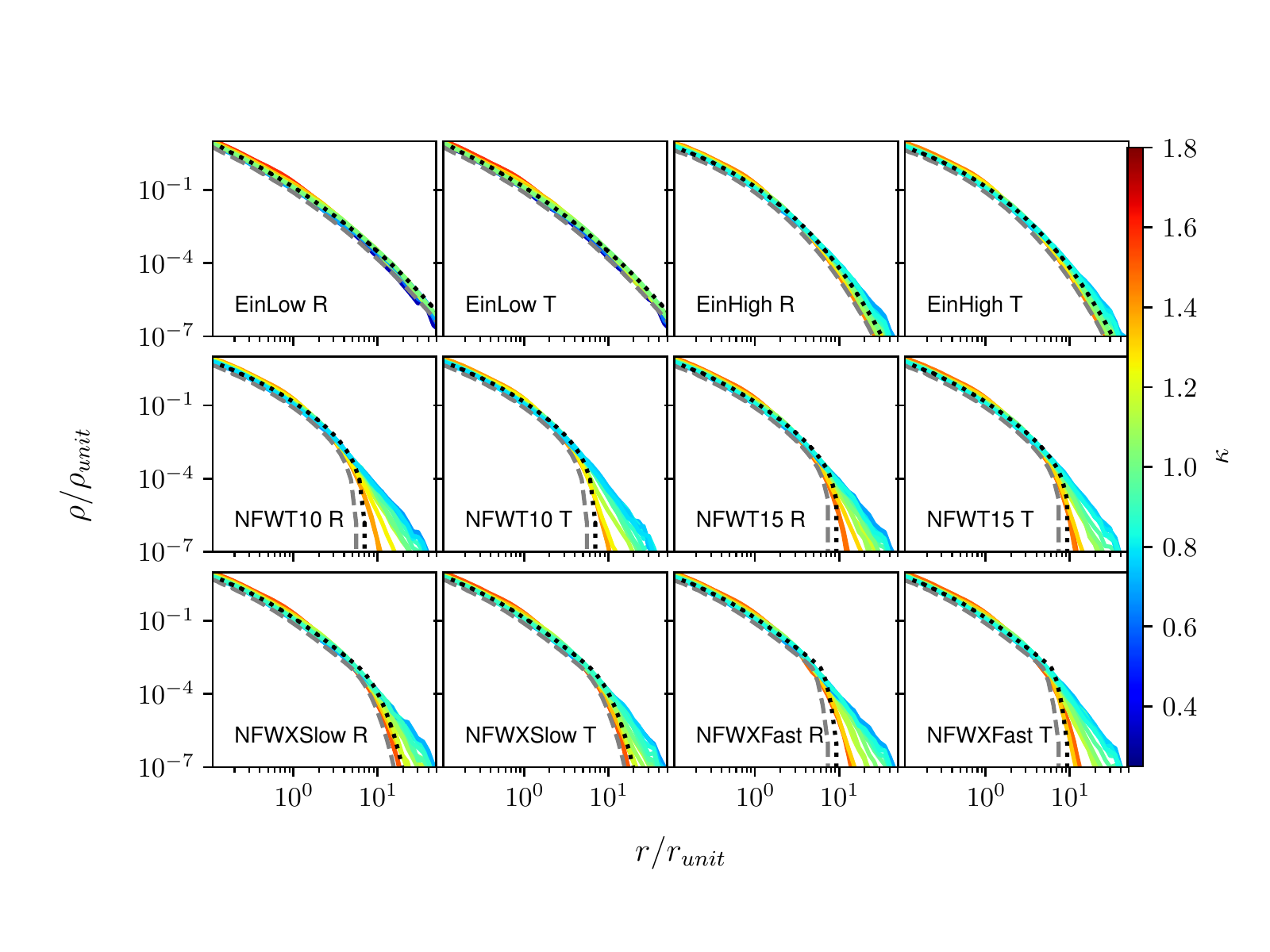}.
	\caption{Density profiles of the halo remnants. The dashed grey lines show the initial halo models, and the dotted black lines are the ICs with the radius rescaled by a factor of $2^{1/3}$, as expected for a self-similar equal-mass merger. The labels indicate the initial halo models, as well as whether the initial velocity was tangential (T) or radial (R). The remnants are coloured by the relative energy change, $\kappa$.}
	\label{fig:ProfRemnantDens}
\end{figure*}

In Fig.~\ref{fig:ProfRemnantDens} we show the final density profiles of the remnants, compared to the initial halo models (dashed grey lines). The density increases at any given radius for all remnants, which is perhaps not surprising given the total mass of the system has doubled. We also show the initial profiles with the radii  scaled by $2^{1/3}$ (dotted black lines); this is the scaling expected for self-similar evolution in an equal-mass merger, since it will conserve the mean density as the mass doubles.

The final remnants have been coloured by the relative energy parameter, $\kappa$. At small radii, the changes in the profile  compared to the scaled ICs are subtle, and the inner slope is roughly conserved, in agreement with previous studies \citep{boylankolchin2004, aceves2006, kazantzidis2006, zemp2008, vass2009}. At large radii, the changes in the density profile are more obvious, as mass has moved outwards, relative to the scaled ICs. (Admittedly, some of this is likely due to the artificially truncated nature of the ICs.) The panels have been divided into orbits with radial (R) or tangential (T) initial velocities; as shown in Paper I, these orbits should produce remnants with different 3D shapes. Interestingly, orbital angular momentum has little effect on the spherically averaged density profile. There does, however, seem to be a systematic change in the density profile with $\kappa$; remnants of low-$\kappa$ encounters have more power-law like profiles (low Einasto alpha parameters), whereas remnants of high-$\kappa$ encounters have more sharply truncated profiles (high Einasto alpha parameters).

%%%%%%FIGURE 2%%%%%%

\begin{figure*}
	\includegraphics[trim={0 1cm 0 0cm},clip]{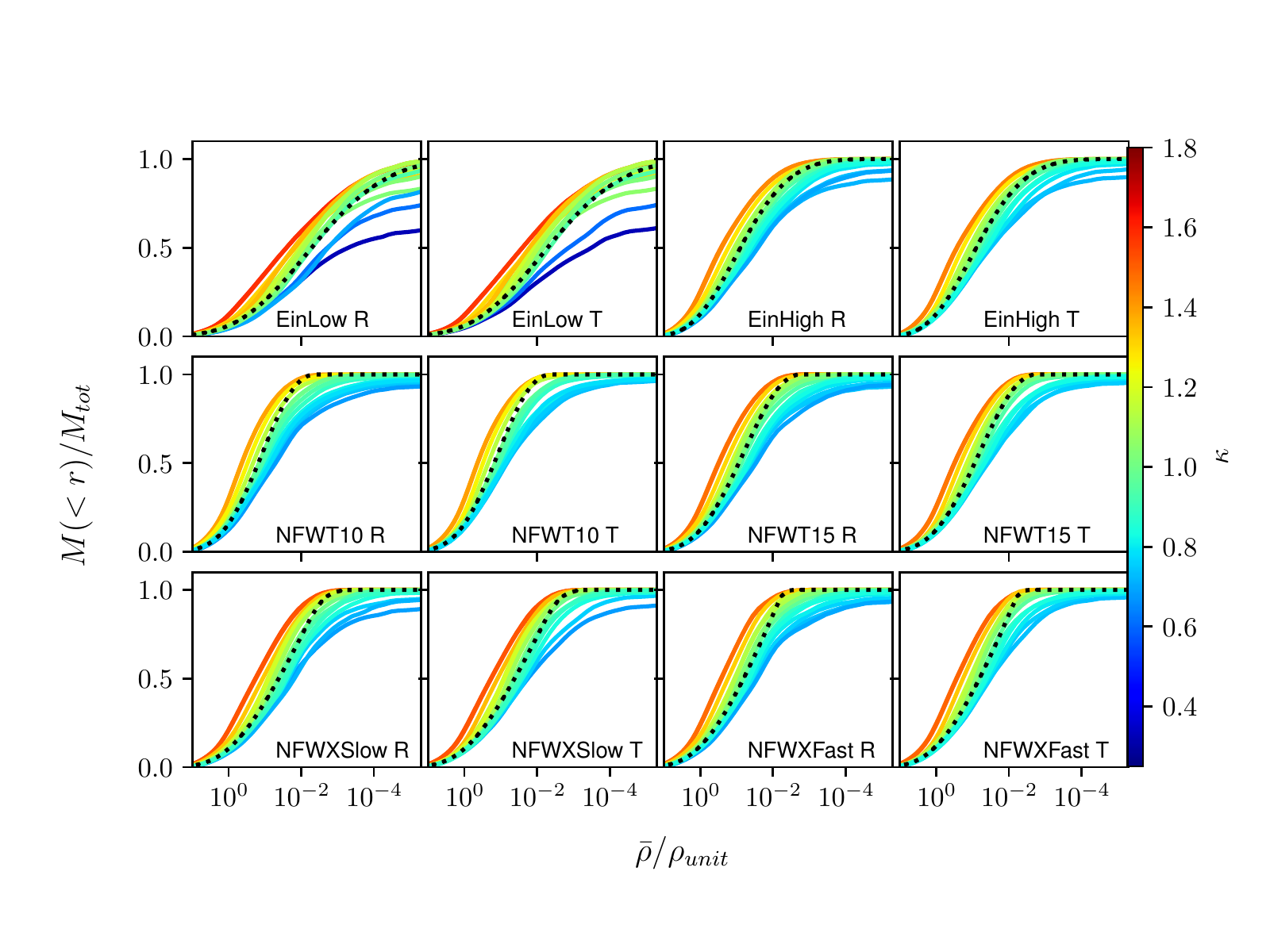}.
	\caption{The enclosed mass fraction within a given radius in the merger remnant, versus the enclosed mean density within that radius. The initial haloes are shown with dotted black lines. The labels indicate the IC model, and whether the initial velocity was purely tangential (T) or purely radial (R). The curves are coloured by the relative energy change $\kappa$, which indicates the change in internal energy relative to scaled ICs. Note that density increases to the left on the horizontal axis, to match the orientation of Fig.~\ref{fig:ProfRemnantDens}, with larger radii to the right.}
	\label{fig:ProfRemnantMass}
\end{figure*}

Since the change in the density profile is difficult to see, particularly at small radii, in Fig.~\ref{fig:ProfRemnantMass} we also plot the enclosed mass fraction within a given radius versus the mean density within that radius. Note that the direction of the $x$-axis has been reversed here, since large densities correspond to small radii. In this plot it is clear that the mass distribution changes even at small radii. The remnants evolve in a monotonic sequence with $\kappa$, with higher $\kappa$ values producing denser remnants at all mass fractions. Again, we find little dependence on the angular momentum of the orbit. Comparing to the ICs, the final remnant curves lie roughly either inside or outside the IC curve, with the divide occurring at the self-similar energy, $\kappa=1$. Sharply truncated profiles, such as NFWXFast, produce  remnants with more diffuse outer regions, even for $\kappa=1$. In a few instances, the remnant curves also cross over the ICs, or over curves of higher or lower $\kappa$, but it is not clear whether this behaviour is robust to changes in resolution. 

%%%%%%FIGURE 3%%%%%%

Overall, it appears that the density profiles of the haloes evolve in a straightforward way that is mainly dependent on $\kappa$. In general, some mass is ejected to large radii, but the central density does not decrease significantly compared to predictions from self-similar scaling. Surprisingly, there is little difference between the remnants produced by radial and tangential encounters. One implication is that the spherically averaged density profile of a remnant is unrelated to its 3D shape, as we showed in Paper I that the shape of the halo {\it does} vary with the angular momentum of the encounter. 

\subsection{Evolution of structural parameters}
\label{subsec:structparam}

Haloes do not typically have sharp boundaries, and therefore their sizes are not well defined. One convenient measure of the size of extended systems is the gravitational radius, which is defined as:
\begin{equation} \label{eq:rg}
r_{\rm g} \equiv \dfrac{GM^2}{|P|} \,\,\, ,
\end{equation}
where $P$ is the potential energy of the system. For a virialized halo,  $\langle v^2 \rangle = GM/r_{\rm g}$. In practice, however, the potential energy is difficult to measure. Therefore, the half-mass radius, $r_{1/2}$, the radius enclosing half the mass of the halo, is a more convenient quantity to characterize halo size. As discussed in \cite{binney}, $r_{1/2}$ is typically proportional to $r_{\rm g}$, and $r_{1/2}/r_{\rm g}$ can range from about $0.4$ to $0.5$ depending on the system's density profile; thus in general we can estimate that $r_{1/2}/r_{\rm g} \approx 0.45$ to good approximation.

\begin{figure*}
	\includegraphics[trim={0cm 2.5cm 0cm 3cm},clip]{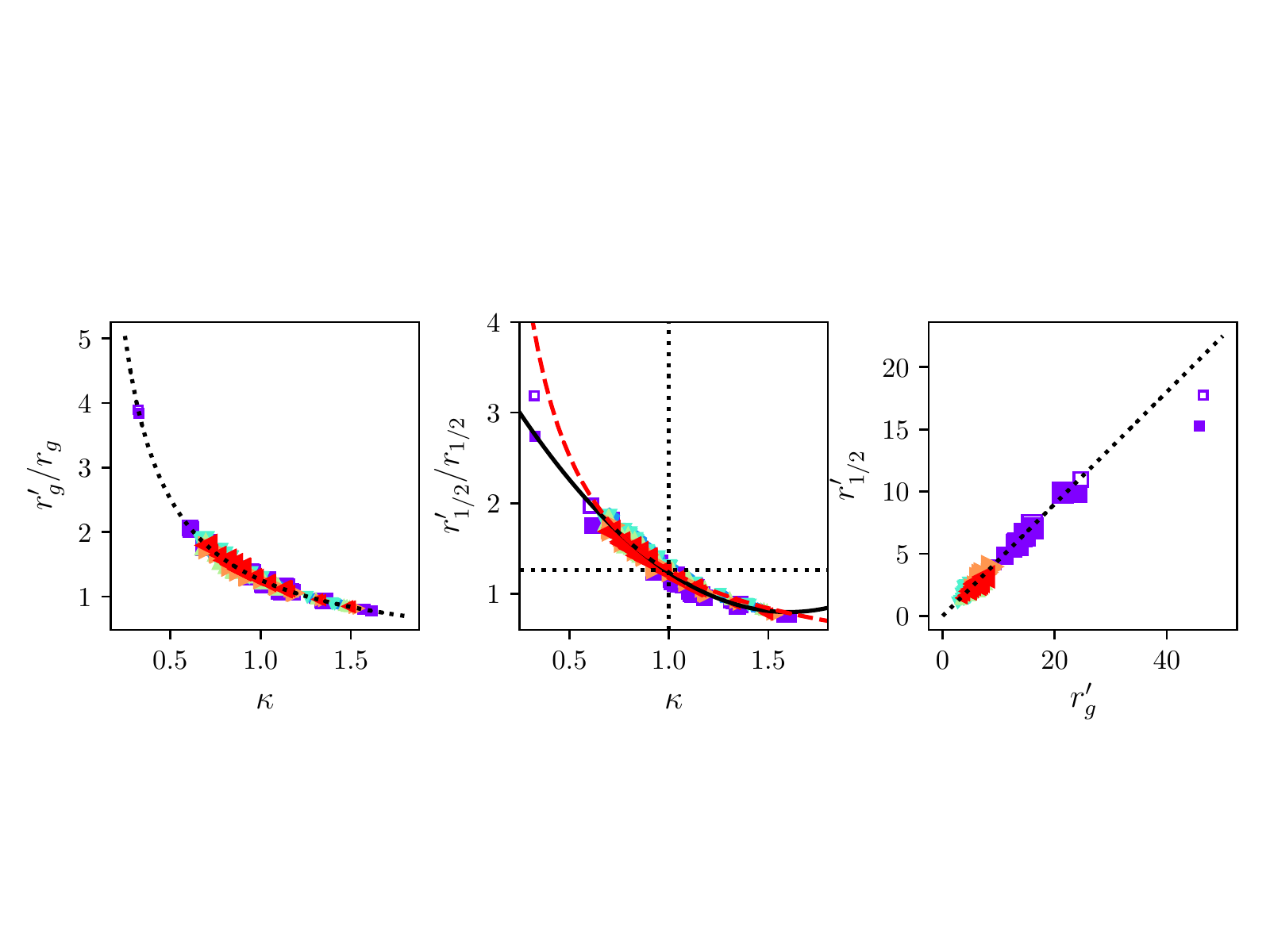}.
	\caption{(Left) The gravitational radius as a function of the relative energy change $\kappa$. The dotted black line is the theoretical expectation, $r_{\rm g} \propto 1/\kappa$. (Middle) The relative change in the half-mass radius, $r_{1/2}'/r_{1/2}$,  as a function of the relative energy change $\kappa$. The black dotted lines show the expected values for self-similar evolution of the density profile. The red dashed curve is the theoretical expectation if $r_{1/2}$ is proportional to $r_{\rm g}$. The solid black line is a fit to the data, $r_{1/2}'/r_{1/2} = 1.2 \kappa^2 - 3.9 \kappa + 3.9$. (Right) The half-mass radius versus the gravitational radius. The dotted line corresponds to $r_{1/2} = 0.45\,r_{\rm g}$. Colours and symbols are as in Fig.~\ref{fig:OrbitalParameters2}.}
	\label{fig:GravitationalAndHalfMass}
\end{figure*}

For the work presented here, $r_{\rm g}$ is a useful quantity since it is closely related to the relative energy change, $\kappa$. For a virialized halo,  $P=2 E_0$, and thus $r_{\rm g} \propto M^2/E_0$. Given our previous definition of the relative energy change,
\begin{equation}
\kappa = \dfrac{E_0'}{E_0} \left( \dfrac{M}{M'}\right)^{5/3}\,,
\end{equation}
and the fact that the initial and final haloes are in virial equilibrium, we have:
\begin{equation} \label{eq:rg_vs_kappa}
\dfrac{r_{\rm g}'}{r_{\rm g}} =  \dfrac{1}{\kappa} \left( \dfrac{M'}{M}\right)^{1/3}= \dfrac{2^{1/3}}{\kappa} \,\,\,.
\end{equation}
We verify this prediction in the left-hand panel of Fig.~\ref{fig:GravitationalAndHalfMass} by calculating the gravitational radius directly from equation~\ref{eq:rg}, and find that it does indeed scale with $1/\kappa$ as expected. This relationship shows that larger values of $\kappa$  (mergers between more highly bound pairs, where the total energy is more negative) result in less diffuse structures.  The middle panel of Fig.~\ref{fig:GravitationalAndHalfMass} shows how the change in half-mass radius varies as a function of $\kappa$. There is a monotonically decreasing relationship between $\kappa$ and $r_{1/2}'/r_{1/2}$, though it shows more scatter than the relationship between $r_{\rm g}'/r_{\rm g}$ and $\kappa$. The dotted lines show where $r_{1/2}'/r_{1/2}=2^{1/3}$, as expected for self-similar evolution. The prediction from equation~\ref{eq:rg_vs_kappa}, assuming $r_{1/2}$ is proportional to $r_{\rm g}$, is also shown (red dashed line). The solid black line is a fit to the data,$r_{1/2}'/r_{1/2} = 1.2 \kappa^2 - 3.9 \kappa + 3.9$; the RMS scatter with respect to this fit is 0.06.

%%%%%%FIGURE 4%%%%%%

We see that the simulation results go through the self-similar expectation at $\kappa=1$. They also agree with the  prediction from equation~\ref{eq:rg_vs_kappa}, except at low $\kappa$ (less bound mergers). To see why this is, in the right-hand panel of Fig.~\ref{fig:GravitationalAndHalfMass} we show $r_{1/2}'$ vs $r_{g}'$. The dotted line corresponds to $r_{1/2} \approx 0.45 r_{\rm g}$. For the most part, the remnants obey the expected relationship, with the exception of two points, corresponding to the simulations with the lowest $\kappa$ values. This deviation from the expected relationship may indicate a departure from self-similarity.

While $r_{\rm g}$ and $r_{1/2}$ are useful for describing how the size of a halo changes in a major merger, concentration measurements typically depend on a characteristic scale radius, $r_{-2}$ and the virial radius, $r_{\rm vir}$, or possibly on the peak of the circular velocity curve, $v_{\rm peak}$ and the circular velocity at the  virial radius, $v_{vir}$. The scale radius, $r_{-2}$ (the radius at which the logarithmic slope of the density profile is $-2$) is difficult to measure accurately, since it requires numerical differentiation; to help with this, we apply a Gaussian smoothing kernel with a $\sigma = 0.1\, \ln{r_{\rm peak}}$ to the derivative of the profile ${\rm d}\ln \rho/{\rm d} \ln r$. We note, however that the resulting value of $r_{-2}$ is somewhat sensitive to the degree of smoothing, and thus we expect some scatter in the results. In the next section, we will also measure $r_{-2}$ using the more commonly employed method of profile fitting, and compare the results to those obtained here in Appendix~\ref{sec:cM}. 

\begin{figure}
	\includegraphics[width=\columnwidth,trim={0cm 5.5cm 8cm 0},clip]{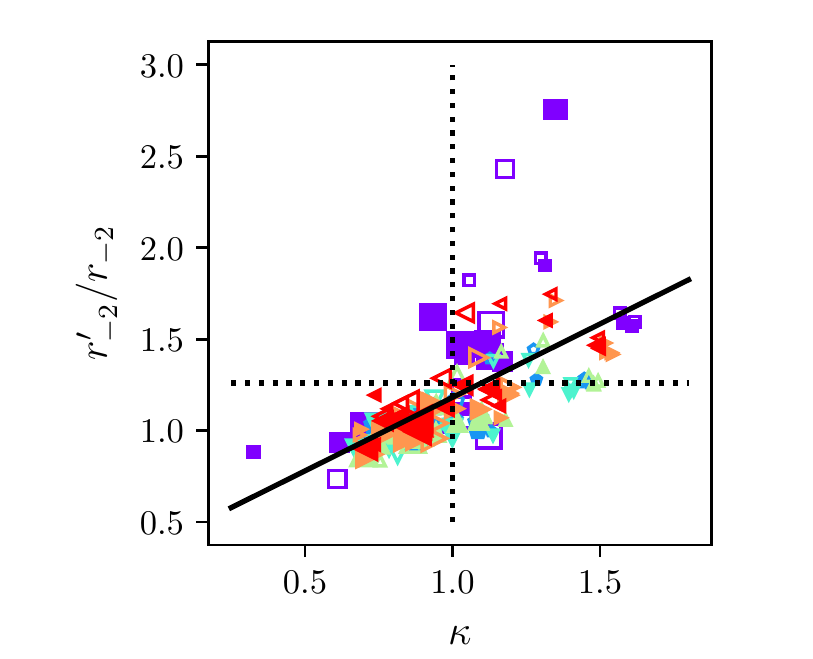}
	\caption{The change in the scale radius, $r_{-2}'/r_{-2}$, as a function of the relative energy change, $\kappa$. The dotted lines are the expectations for self-similar evolution of the density profile. Colours and symbols are as in Fig.~\ref{fig:OrbitalParameters2}. The solid black line is a fit to the data, $r_{-2}'/r_{-2} = 0.80 \kappa + 0.38$.}
	\label{fig:r2_vs_Energy}
\end{figure}

Despite the greater scatter in the measurements, Fig.~\ref{fig:r2_vs_Energy} shows a clear trend in the scale radius with $\kappa$. Unlike $r_{\rm g}$  or $r_{1/2}$, however, changes in $r_{-2}$ generally {\it increase} with increasing $\kappa$, matching the self-similar prediction at $\kappa=1$ (black dotted lines). The solid black line is a fit to the simulation results, $r_{-2}'/r_{-2} = 0.80 \kappa + 0.38$. This fit has an RMS scatter of 0.3, although the scatter is smaller for $\kappa <1$ and much larger for $\kappa > 1$. Interestingly, the lower $\kappa$ mergers seem to leave the scale radius, $r_{-2}$ unchanged, or even decrease it slightly. This is surprising, since these correspond to more `violent' encounters with more initial kinetic energy. Naively, these mergers might be expected to heat up the central parts of the haloes, producing a remnant with lower central density and a larger scale radius. Fig.~\ref{fig:ProfRemnantMass} shows that while the profile of the remnant has expanded relative to the self-similar prediction, the resulting density change is much larger at large radii than at small radii. As a result, the scale radius does not necessarily increase, and can even decrease.

%%%%%%FIGURE 5%%%%%%

\begin{figure*}
	\includegraphics[trim={0 1cm 0 0cm},clip]{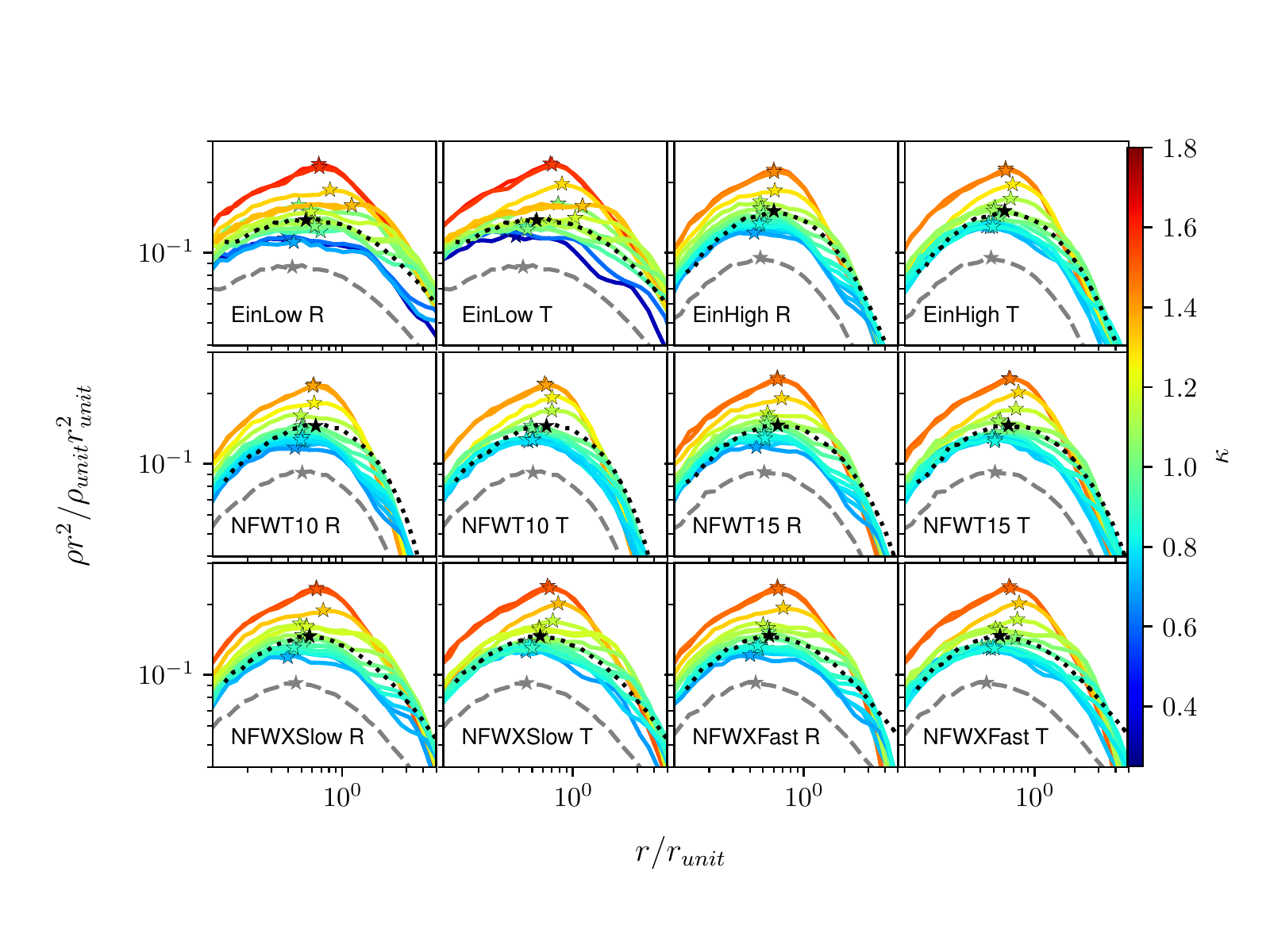}.
	\caption{
		$\rho r^2$ versus radius for the halo remnants, coloured by the relative energy change, $\kappa$. Scale radii, $r_{-2}$, are indicated with stars. The dashed grey lines show the initial halo models, and the dotted black lines are the ICs with the radius rescaled by a factor of $2^{1/3}$, as expected for a self-similar equal-mass merger.}
	\label{fig:ProfRemnantDensR2}
\end{figure*}

To understand why the scale radius increases with $\kappa$, while other radii decrease, in Fig.~\ref{fig:ProfRemnantDensR2} we plot $\rho r^2$ versus radius. The peak of the curve indicates the radius at which $r=r_{-2}$. We note several interesting features. First, the profiles become more sharply peaked (i.e. they have higher Einasto alpha parameters) for larger values of $\kappa$. Second, the curves are not completely smooth, but have several small variations relative to the scaled ICs. These might appear to be noise in the density profiles, but the fact that they reoccur at similar radii in simulations with different $\kappa$ values (e.g.~in the NFWT15 R or NFWXSlow R panels) suggests that they are real features in the remnants. Similar variations can be seen in merger remnants from e.g. \cite{mcmillan2007}. For very flat profiles, such as the EinLow simulations, these features make the scale radius hard to measure and add scatter to the measured values. 

Third, if we consider the difference between the scaled ICs (dotted black lines) and the final remnants, we see that the largest changes in density are at, or past, the scale radius. At small radii, the density changes less, and only ever drops $\sim$10\%\ below the corresponding value for the scaled ICs. The net effect of the density changes being larger at large radii is that while low-$\kappa$ simulations produce more extended profiles (see Fig.~\ref{fig:ProfRemnantDens}), they actually have {\it smaller} scale radii. Conversely, very bound (high $\kappa$) remnants have large scale radii, but are then truncated abruptly beyond this radius.

%%%%%%FIGURE 6%%%%%%

\begin{figure}
	\includegraphics[trim={3.5cm 0cm 4.5cm 0cm},clip,width=\columnwidth]{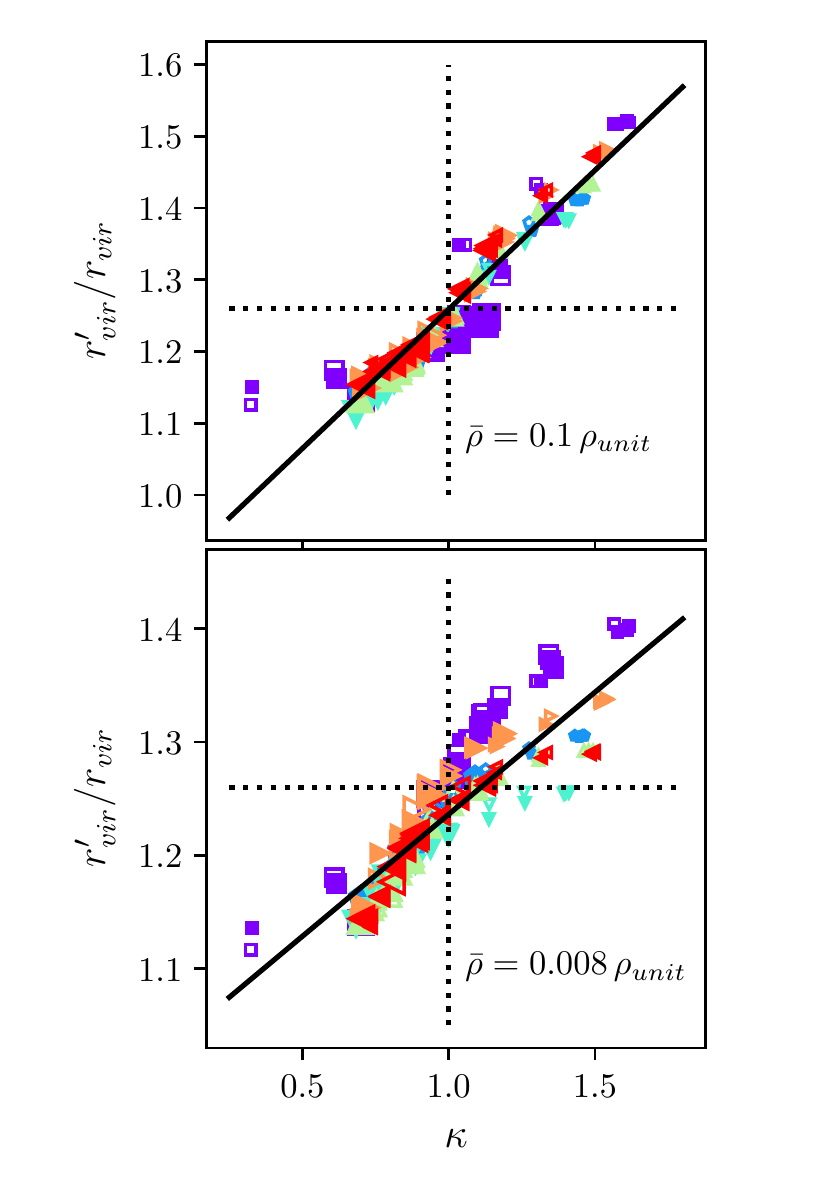}.
	\caption{The change in virial radius, $r_{\rm vir}'/r_{\rm vir}$, as a function of the relative energy change, $\kappa$. The two virial radii are defined in terms of enclosed mean density, as described in the text, for densities of $\bar{\rho} = 0.1\,\rho_{\rm unit}$ (top) and $0.008\,\rho_{\rm unit}$ (bottom). The dotted lines show the expected values for self-similar scaling of the density profile. Colours and symbols are as in Fig.~\ref{fig:OrbitalParameters2}. The solid lines are fits to the data, $r_{\rm vir}'/r_{\rm vir} = 0.39 \kappa + 0.87$ (top) and $r_{\rm vir}'/r_{\rm vir} = 0.22 \kappa + 1.02$ (bottom).}
	\label{fig:rvir}
\end{figure}

Next, we consider the virial radius, $r_{\rm vir}$. For cosmological haloes, this is formally the radius within which the system is in virial equilibrium. Strictly speaking, this definition only applies to systems that are accreting continuously from the surrounding density field; there is no clear analogue of this quantity in our isolated simulations. In practice, however, the virial radius in cosmological simulations is usually defined as the radius within which the mean density of the halo exceeds a reference background density $\rho_{\rm ref}$ by a factor $\Delta$. A common choice is $\rho_{\rm ref} = \rho_{\rm c}$, the critical density, and $\Delta = 200$ \citep{navarro1996, navarro1997}. By analogy, we will consider two `virial radii' that enclose mean densities $\bar{\rho} = 0.1\,\rho_{\rm unit}$  and $0.008\,\rho_{\rm unit}$. Given our choice of units, for an NFW profile these would correspond to 3 and 10 times the scale radius, respectively (or concentration parameters of $c=3$ and $c=10$), in our ICs. (Given these definitions, we also note that the `virial mass' within each of these radii will scale as $M_{\rm vir} \propto r_{\rm vir}^{3}$.) Fig.~\ref{fig:rvir} shows the change in these radii as a function of $\kappa$. The solid lines are fits to the data, $r_{\rm vir}'/r_{\rm vir} = 0.39 \kappa + 0.87$ (top) and $r_{\rm vir}'/r_{\rm vir} = 0.22 \kappa + 1.02$ (bottom), with an RMS scatter of 0.03 for both. Overall, the relationship between $r_{\rm vir}$ and $\kappa$ is monotonic, and fairly tight, with some possible dependence on the initial profile. As with the scale radius, the change in the virial radius {\it increases} with $\kappa$. Thus, while the remnants produced by more energetic encounters are larger and more diffuse, a virial radius defined in terms of enclosed density is actually smaller for these systems. We note, however, that the slopes of the virial radius--$\kappa$ relations (0.39 and 0.22, for the two density thresholds) are both shallower than the relation for the scale radius (0.8). Thus, although the density change relative to self-similar scaling is larger in the outer parts of the remnant (as shown in Fig.~\ref{fig:ProfRemnantMass}), the relative change in size is smaller at larger radii than at smaller radii.

%%%%%%FIGURE 7%%%%%%

\begin{figure}
	\includegraphics[width=\columnwidth,trim={3.5cm 0cm 4.5cm 0cm},clip]{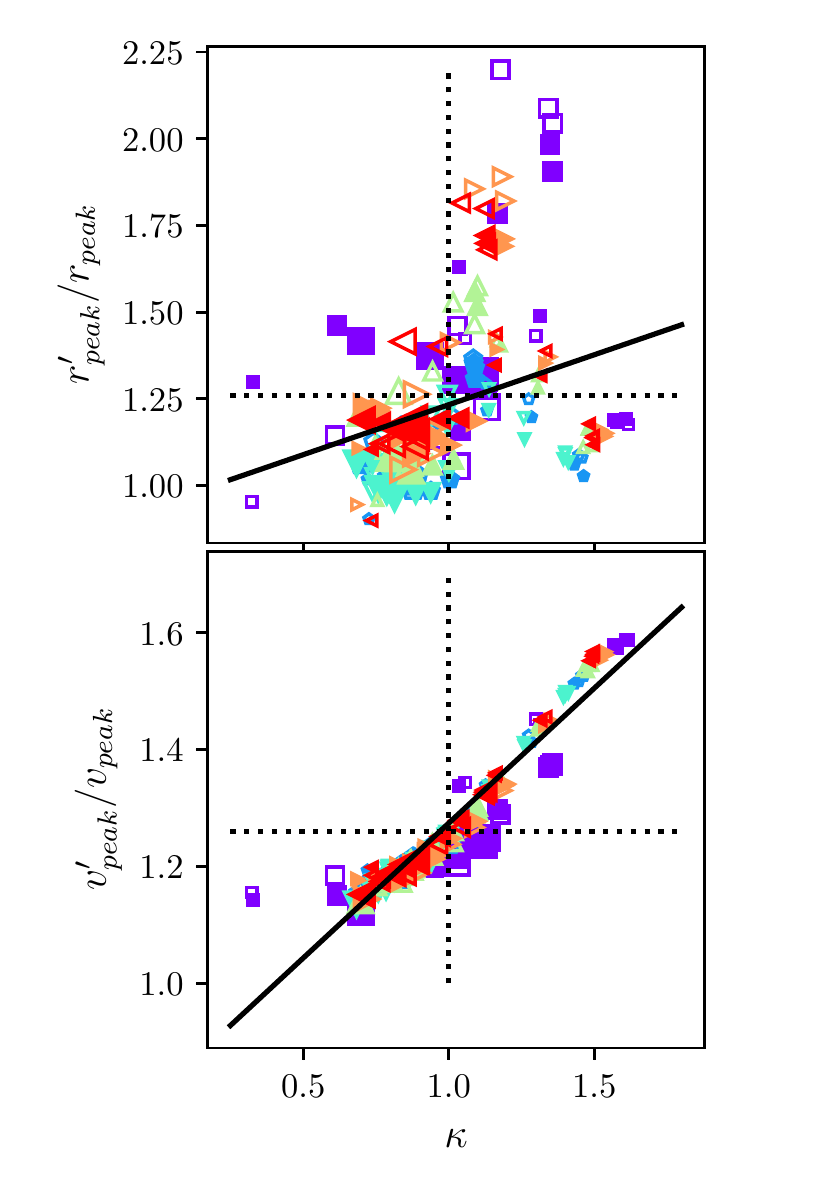}.
	\caption{Changes in $r_{\rm peak}$ (top) and $v_{\rm peak}$ (bottom) as a function of the relative energy change, $\kappa$. 
	The dotted lines show the values for self-similar scaling. Colours and symbols are as in Fig.~\ref{fig:OrbitalParameters2}. The solid black lines are fits to the data, $r_{\rm peak}'/r_{\rm peak} = 0.29 \kappa + 0.95$ (top) and $v_{\rm peak}'/v_{\rm peak} = 0.46 \kappa + 0.81$ (bottom).}
	\label{fig:vpeak}
\end{figure}

Finally, we examine how the peak circular velocity, $v_{\rm peak}$, and the corresponding radius, $r_{\rm peak}$, vary from the ICs to the final remnant. Fig.~\ref{fig:vpeak} shows the relative changes in $r_{\rm peak}$ (top) and $v_{\rm peak}$ (bottom) as a function of the relative energy change, $\kappa$. The solid black lines are fits to the data, $r_{\rm peak}'/r_{\rm peak} = 0.29 \kappa + 0.95$ (top) and $v_{\rm peak}'/v_{\rm peak} = 0.46 \kappa + 0.81$ (bottom); the RMS scatter with respect to these are 0.2 and 0.04, respectively. The relative change in peak velocity increases approximately linearly with $\kappa$. The relationship between $r_{\rm peak}$ and $\kappa$ is more complicated; here the points follow two trends. The low-energy simulations generally produce remnants with peak radii close to those of the ICs, that is to say smaller than the self-similar expectation, while for the high-energy simulations $r_{\rm peak}$ generally increases linearly with $\kappa$, passing through the self-similar value at $\kappa = 1$. There are a number of exceptions to these trends, however, notably the simulations with EinLow profiles.

%%%%%%FIGURE 8%%%%%%

In Paper I, we found that the size of haloes, measured as the average particle distance from the halo centre, decreased with $\kappa$ as $\kappa^{-5}$. The results in this paper indicate that while the radii  $r_{\rm g}$ and $r_{1/2}$ also decrease (albeit as $\kappa^{-1}$), the characteristic radii $r_{-2}$, $r_{\rm vir}$ and $r_{\rm peak}$  all {\it increase} roughly linearly with $\kappa$. Clearly the changes in the profile are complicated and non-self-similar, suggesting that the behaviour of the concentration parameter may be complicated. Finally, since there is a considerable literature investigating how the inner slope of halo profiles evolve in mergers, we have also included a discussion of this point in Appendix~\ref{sec:innerslope}.

\section{Profile Fitting} \label{sec:ProfFit}

In the previous section, we examined how the overall mass distribution and the characteristic radii evolve in major mergers. In this section, we fit analytic NFW and Einasto profiles to the remnants, and discuss how well these analytic forms describe the remnants.

\subsection{Methods}
\label{subsec:4.1}

To determine the NFW parameters ($r_{\rm s}, \rho_0$) or the Einasto parameters ($r_{-2}, \rho_{-2}, \alpha_{\rm E} $), we fit the circular velocity profiles using a $\chi^2$-minimization procedure. Though previous studies generally fit the density profile directly \citep[e.g.][]{neto2007, duffy2008,dutton2014,meneghetti2014}, fitting the circular velocity is less susceptible to noise, as discussed in \cite{veraciro2013}.

The data were binned in logarithmic radial bins, with each bin centred on $\ln r_i$. As in \cite{veraciro2013,veraciro2014}, we fit the profiles by minimizing
\begin{equation}
	\chi^2 =\dfrac{1}{N_{bins}} \sum_{i=1}^{N_{bins}} (\ln v_{\rm c}^2 - \ln v_{{\rm c},i}^2)^2 \,\,\, ,
\end{equation}
where $N_{bins}$ is the number of bins, $v_c$ is the circular velocity of the fitted profile at radius $r_i$, and $v_{c,i}$ is the circular velocity of the simulated halo at $r_i$.  The fit was performed between three times the softening length and the radius at which the mean enclosed density was $\bar{\rho}=0.01 \, \rho_{\rm unit}$. Though we present both NFW and Einasto fits, it has been shown that concentrations from Einasto fits are more robust to variations in fitting details, such as the radial range \citep{gao2008}.

\subsection{Fits to individual remnants}
\label{subsec:4.2}

\begin{figure*}
	\includegraphics[scale=1.1, trim={0 0 0 0},clip]{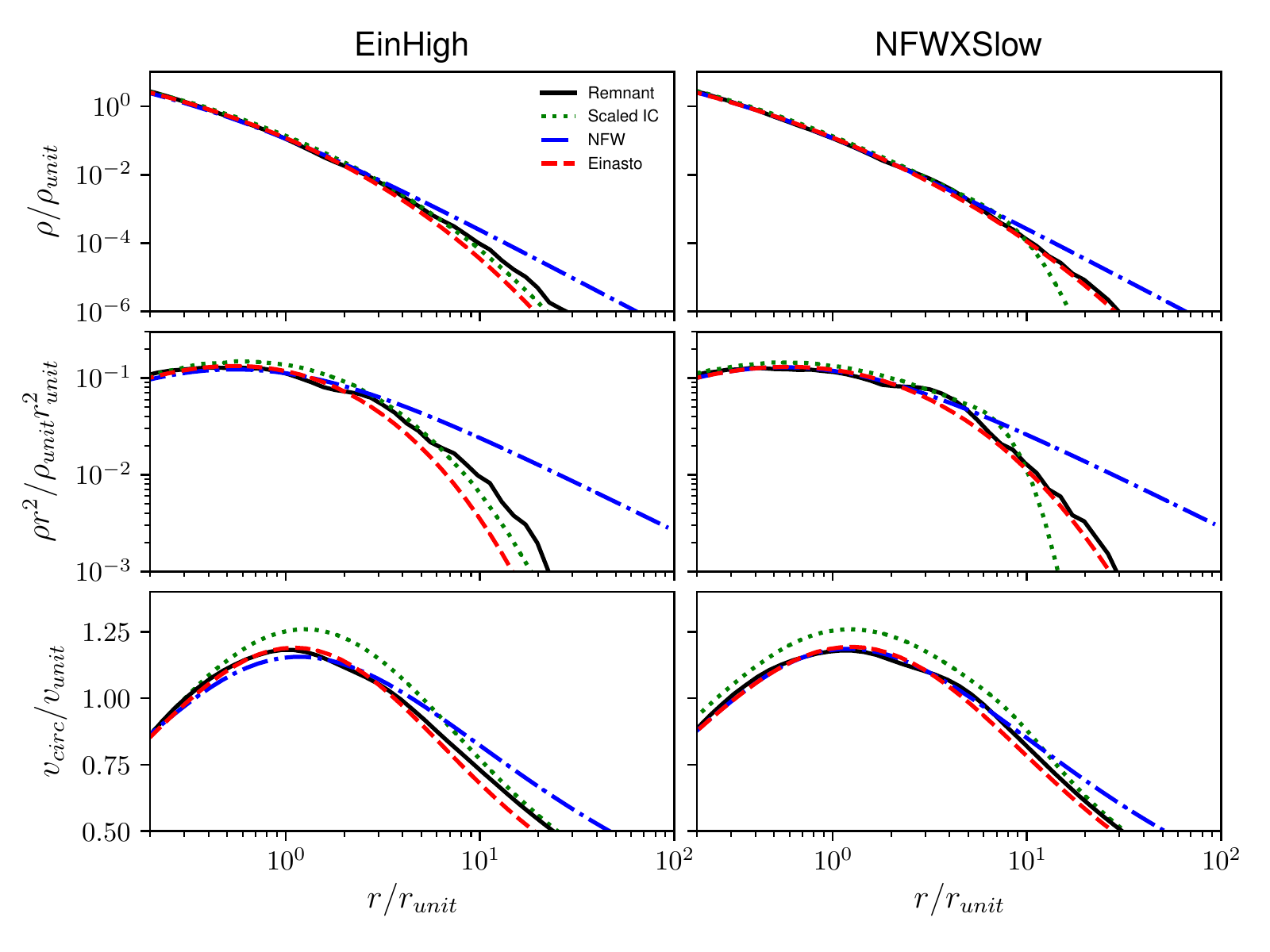}.
	\caption{Density profile, $\rho$ (top), $\rho r^2$ (middle), and the circular velocity curve $v_{\rm circ}$ (bottom) for two example remnants. The initial ICs were either EinHigh (left) or NFWXSlow (right), and merged on a radial orbit at a separation of $10\,r_{\rm peak}$ with an initial velocity of $0.8\,v_{\rm esc}$. The remnant and ICs are shown with solid black and dotted green lines, respectively. The ICs have been scaled as expected for self-similar evolution in an equal-mass merger; i.e., the radii and velocities have been scaled by $2^{1/3}$. The best-fitting NFW (blue dash-dotted line) and Einasto (red dashed line) profiles are also shown.}
	\label{fig:SelfSim}
\end{figure*}

 Fig.~\ref{fig:SelfSim} shows the merger remnants of radial mergers with EinHigh (left) and NFWXSlow ICs (right). Einasto profiles provide a much better description of the remnants in both cases. This may not be surprising, given that the ICs had either Einasto or truncated profiles. Even for the Einasto profile, however, there is a change in the $\alpha_{\rm E}$ parameter. These results demonstrate that the profiles of the remnants are not, in general, NFW, nor are they self-similar to the ICs.

%%%%%%FIGURE 9%%%%%%

\begin{figure}
	\includegraphics[trim={3.5cm 0cm 5cm 0cm},clip]{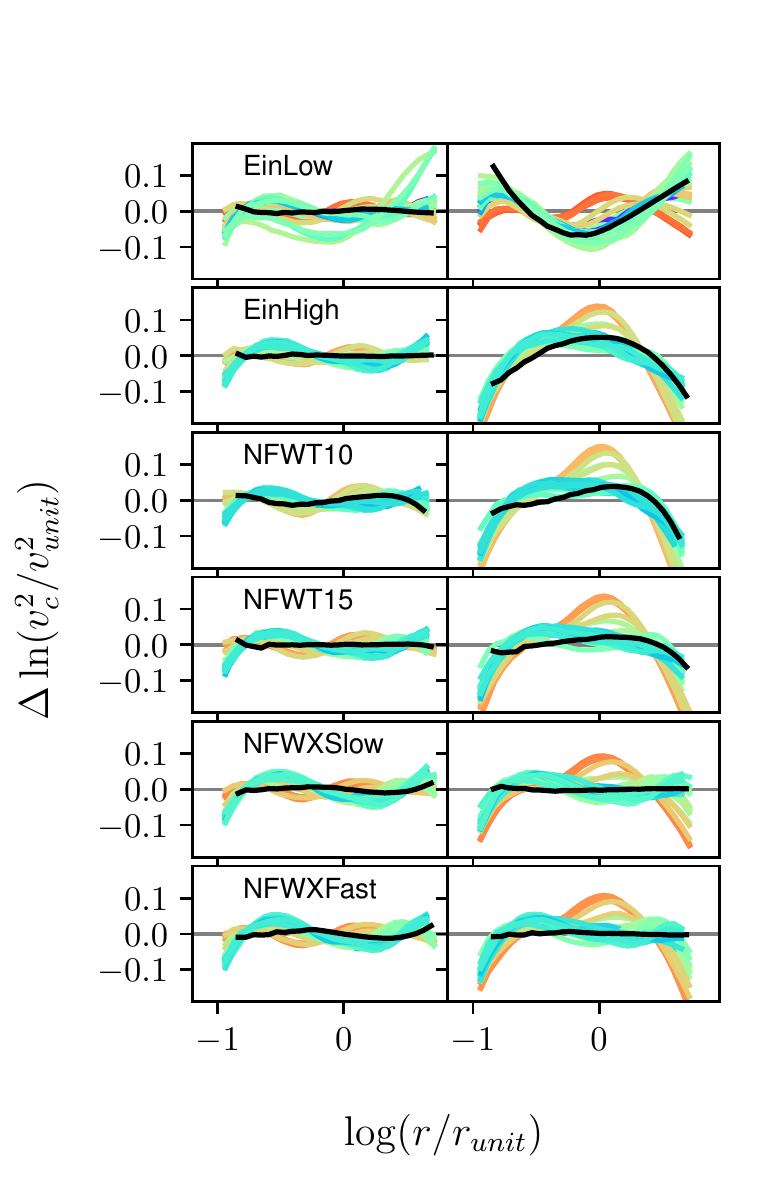}.
	\caption{Residuals in the Einasto (left) and NFW (right) fits to the individual remnants. The residuals are measured as the difference in $\ln v_c^2$ between the remnant and the fits. The halo remnants are coloured by the relative energy change, $\kappa$. Residuals in the fits to the ICs are shown in black.}
	\label{fig:Residuals}
\end{figure}

To assess how well the haloes are fit by NFW and Einasto profiles, we show the residuals in Fig.~\ref{fig:Residuals}, in the range that the velocity profiles were fit. The remnants are coloured by their relative energy parameter (where redder lines correspond to higher $\kappa$ values), and fits to the ICs are shown in black. Overall, the residual of the remnant fits are more consistent with zero for the Einasto fits when compared to the NFW fits. The residuals look very similar for a given relative energy for all the ICs, with the  exception of the EinLow simulations. The high-$\kappa$ simulations show different trends in their residuals compared to the low-$\kappa$ simulations. Comparing the ICs to the profile fits, we can see that the Einasto profiles are well fit with Einasto profiles, and NFWX profiles by NFW profiles, as expected. Interestingly, the NFWT ICs have very low residuals when fit with an Einasto profile; since these profiles resemble tidally stripped systems \citep{drakos2017}, this shows that Einasto profiles are a reasonable description of tidally stripped NFW profiles.

\begin{figure}
	\includegraphics[trim={2.2cm 0 3.8cm 3cm},clip,width=\columnwidth]{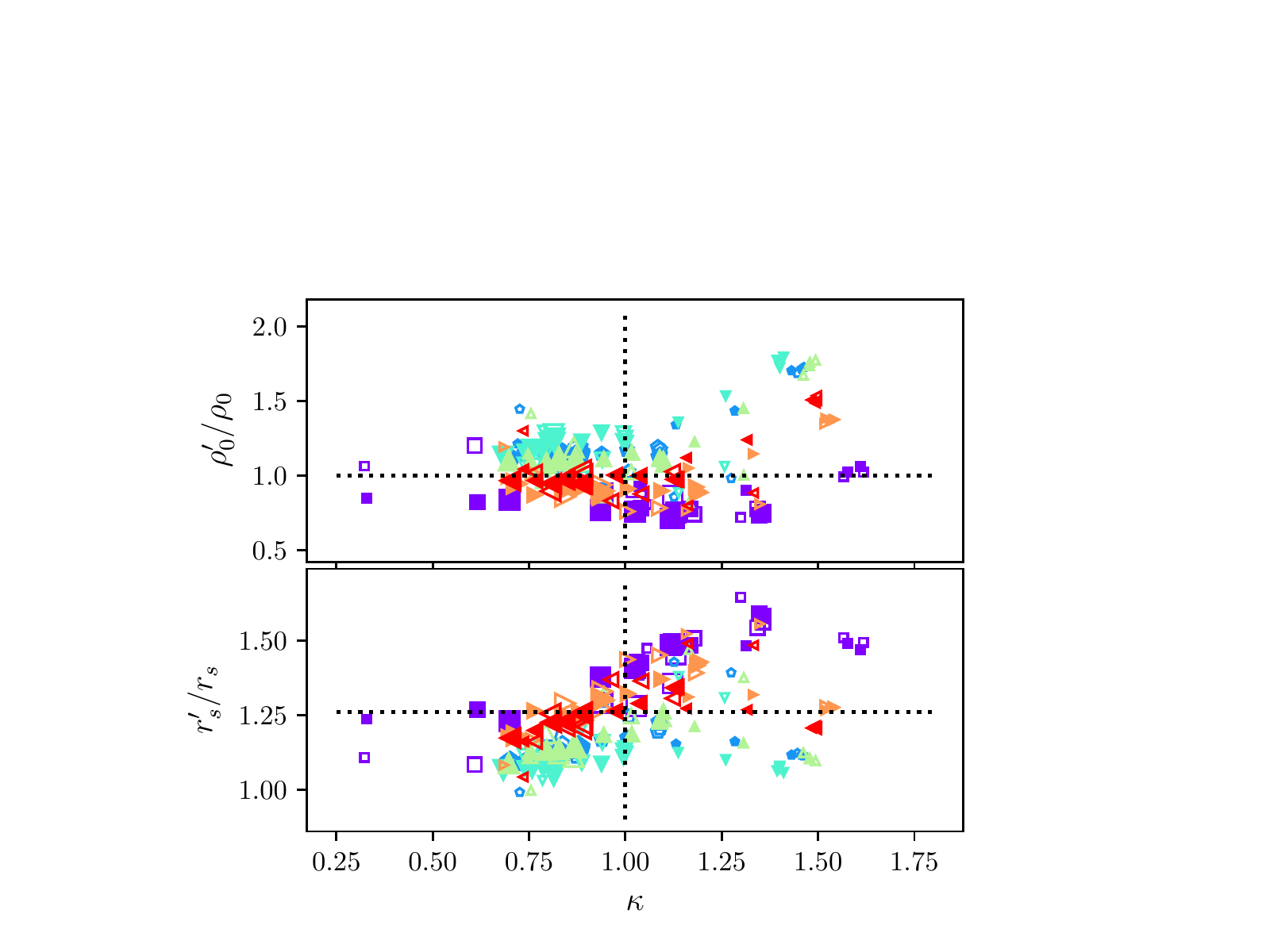}.
	\caption{The change in the best-fitting NFW parameters, $r_{-2}$ and $\rho_{0}$, as a function of the relative energy change, $\kappa$. The dotted lines are the expectations for self-similar evolution of the density profile. Colours and symbols are as in Fig.~\ref{fig:OrbitalParameters2}.}
	\label{fig:NFWparams}
\end{figure}

Fig.~\ref{fig:NFWparams} shows the changes in the NFW parameters, $\rho_0$ and $r_{\rm s}$, as a function of $\kappa$. The density parameter, $\rho_0$ has little dependence on $\kappa$, with the exception of the high-$\kappa$ simulations. The change in the scale radius, $r_{\rm s}'/r_{\rm s}$, generally increases with increasing $\kappa$, as was found previously from direct fitting (cf.~Fig.~\ref{fig:r2_vs_Energy}). Both parameters roughly match the self-similar prediction for $\kappa = 1$ (where the dotted lines intersect). At high energies (low-$\kappa$), it is generally found that $r_{\rm s}'/r_{\rm s}<2^{1/3}$, i.e. $r_{\rm s}$ is smaller than expected from self-similar evolution. We also note that there appears to be some systematic dependence on the initial halo model in both sets of results (as indicated by the point colour).

The lowest energy simulations (high-$\kappa$ values) appear to behave slightly differently from the other simulations; the density increases more than for the other simulations, while the trend in $r_{\rm s}$ with $\kappa$ is reversed at these energies. This change in behaviour may indicate that these remnants are no longer well fit by NFW profiles. In what follows, we will generally distinguish three groups of simulations with somewhat distinct behaviour: the EinLow simulations, the lowest energy (highest $\kappa$) simulations, and the rest of the simulations, which follow a simpler trend. 

%%%%%%FIGURE 10%%%%%%

\begin{figure}
	\includegraphics[trim={2.2cm 0 3.8cm 0cm},clip,width=\columnwidth]{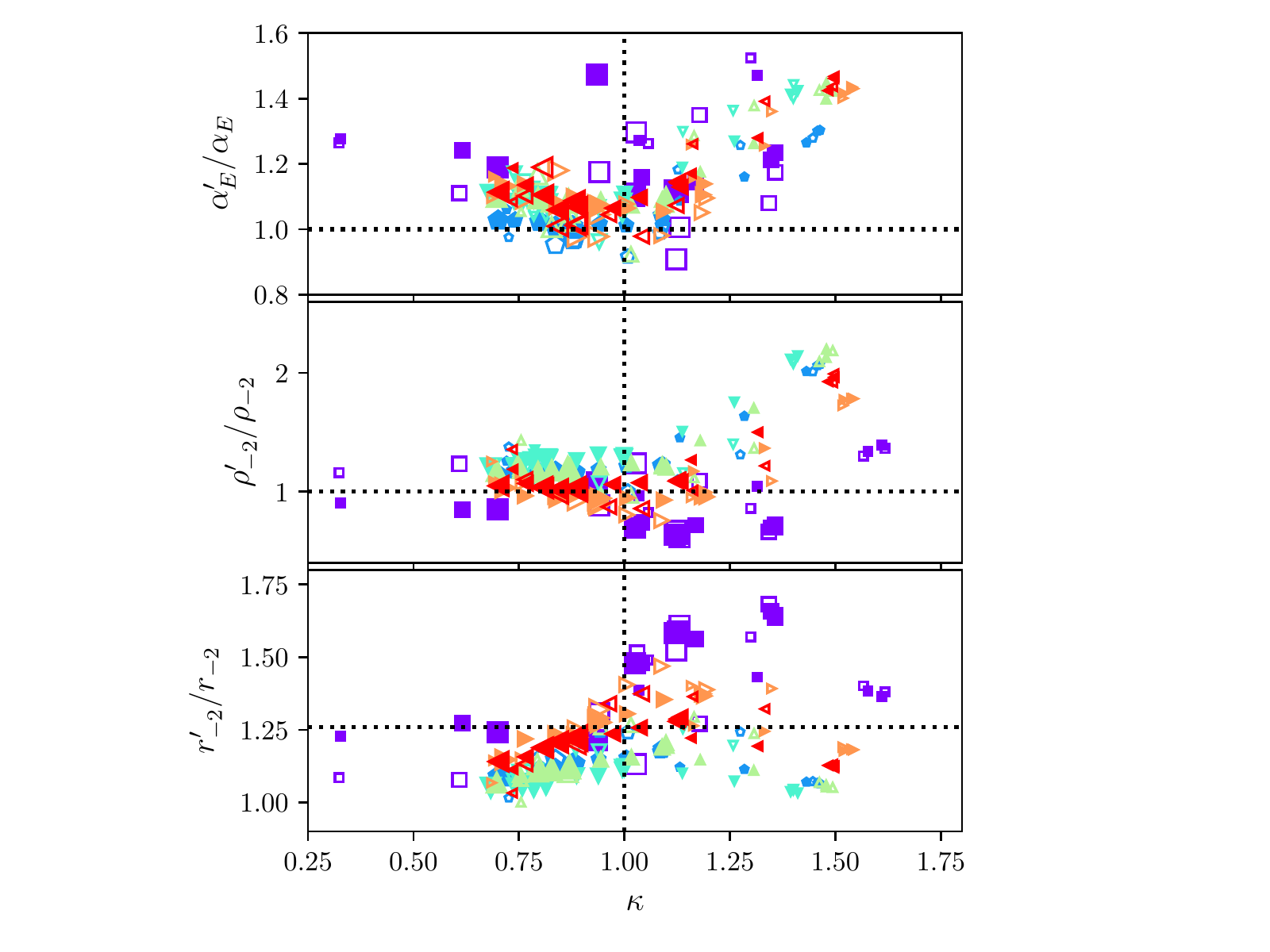}.
	\caption{
	The change in the best-fitting Einasto parameters, $\alpha_{\rm E}$, $\rho_{-2}$ and $r_{-2}$ as a function of the relative energy change, $\kappa$.  The dotted lines are the expectations for self-similar evolution of the density profile. Colours and symbols are as in Fig.~\ref{fig:OrbitalParameters2}. }
	\label{fig:Einparams}
\end{figure}

Similarly, the variations in the three Einasto parameters $\alpha_{\rm E}$, $\rho_{-2}$ and $r_{-2}$ are shown in Fig.~\ref{fig:Einparams}. The variations in $\alpha_{\rm E}$ and $\rho_{-2}$  are roughly independent of $\kappa$, although, as with the NFW parameters, the high-$\kappa$ simulations differ considerably from the others, producing remnants with higher $\alpha_{\rm E}$ and $\rho_{-2}$ values. Variations in $r_{-2}$ have a stronger dependence on $\kappa$; $r_{-2}$ generally increases more for larger values of $\kappa$, though once again the high-$\kappa$ simulations deviate from the main pattern. As with the NFW profile, for all three Einasto parameters, the main trends only roughly match the prediction for self-similar scaling when $\kappa = 1$. This could be because the profile is not perfectly described by an Einasto model, especially in the outer regions, and the parameter values are very sensitive to the exact profile fit. 
%Our model predictions from Equations~\ref{eq:EinParamFits1}--\ref{eq:EinParamFits3}  are shown as dashed lines, for two values of $E_0$. These seem reasonable for $\kappa$ = 0.7--1.5; outside this range our sampling of parameter space is sparse, so the behaviour is less clear.

%%%%%%FIGURE 11%%%%%%

In summary, we find that the remnants of equal-mass binary mergers are not self-similar to their ICs, but are fairly well described by Einasto profiles. Changes in the halo profile depend mainly on the relative change in internal energy, $\kappa$, though there is also some dependence on the initial model of the halo.  Interestingly, the Einasto shape parameter $\alpha_{\rm E}$ typically increases in mergers. High peaks in the density field, and/or rapidly forming haloes, have higher values of  $\alpha_{\rm E}$ \citep{gao2008, ogiya2016}; it is possible that mergers have a role in producing these profiles.

\subsection{Analytic model of profile changes}
\label{subsec:4.3}

Ultimately, we would like a synthetic prediction for how the profile of the remnant will differ from that of the ICs. We can specify the relationship fully by determining the changes in the three Einasto parameters, $\alpha_{\rm E}$, $\rho_{-2}$, and $r_{-2}$, as a function of the ICs and/or merger parameters. Clearly, we need three independent equations to predict the changes in these three parameters. Ideally, we would use the equations for $r_{1/2}$, $v_{\rm peak}$, and $r_{\rm vir}$ derived in Section~\ref{subsec:structparam}, since they have little dependence on the initial halo model. In practice, however, we have found that this method does not work well, as the Einasto parameters are very sensitive to small variations in these equations. Instead, we determine the three Einasto parameters from profile fits to the merger remnants, and fit directly for the dependence on  $\kappa$ and $E_0$. We assume that the dependence is at most quadratic in both variables. The resulting fits are:

\begin{subequations} \label{eq:EinParamFits}
\begin{align}
 \label{eq:EinParamFits1}
\alpha_{\rm E}'&=  
(0.03 \kappa^2 -0.06 \kappa +0.06) \times \\ \notag &\,\,\,\,\,\,\,\,\,\,
\left(0.37 \dfrac{E_0^2}{E_{\rm unit}^2}+4.72 \dfrac{E_0}{E_{\rm unit}} +13.2 \right) \\
  \label{eq:EinParamFits2}
\dfrac{\rho_{-2}'}{\rho_{\rm unit} }&= 
(0.96\kappa^2-1.75\kappa + 1.19)\times \\ \notag &\,\,\,\,\,\,\,\,\,\,
\left(0.46 \dfrac{E_0^2}{E_{\rm unit}^2} + 1.87 \dfrac{E_0}{E_{\rm unit}} +2.68\right) \\
 \label{eq:EinParamFits3}
\dfrac{r_{-2}'}{r_{\rm peak}}&= 
(-0.11\kappa^2 +0.25 \kappa +0.06)\times \\ \notag &\,\,\,\,\,\,\,\,\,\,
\left(-0.75 \dfrac{E_0^2}{E_{\rm unit}^2} - 2.79 \dfrac{E_0}{E_{\rm unit}}+ 0.57\right) \,\,\,.
\end{align}
\end{subequations}

\begin{figure}
	\includegraphics[width=\columnwidth,trim={2.5cm 0cm 3.5cm 0cm},clip]{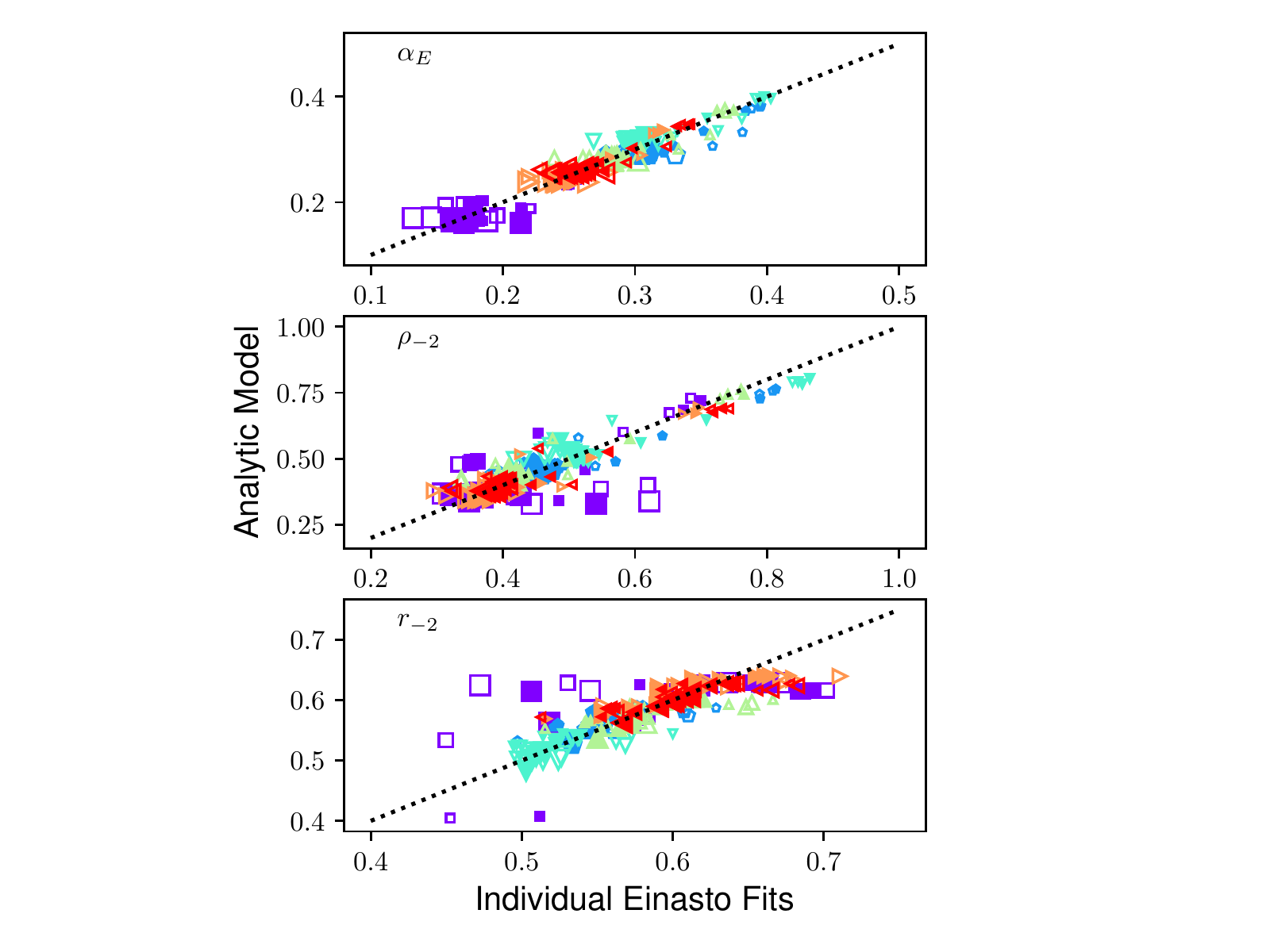}.
	\caption{Einasto parameters $\alpha_{\rm E}$, $\rho_{-2}$, and $r_{-2}$, predicted by our fits (equations~\eqref{eq:EinParamFits1}-\eqref{eq:EinParamFits3}), versus the parameters values measured from the best-fitting profile to each remnant halo.
		}
	\label{fig:ModelFits}
\end{figure}

Fig.~\ref{fig:ModelFits} shows how the predicted Einasto parameters compare to those derived from profile fits to the individual remnants. The RMS scatter with respect to the fit is 0.02, 0.06, and 0.03 for the top, middle, and bottom panels, respectively. In general, the scatter is fairly small, except for the EinLow profiles (squares). 
Remnants from EinLow ICs have low-$\alpha_{\rm E}$ parameters, and most have fairly flat logarithmic slopes to their density profile, as illustrated in Fig.~\ref{fig:ProfRemnantDensR2}; these are the most massive, extended profiles with a large fraction of their mass at large radii, possibly explaining the deviations from the general trend. 
Overall, we conclude that equations~\eqref{eq:EinParamFits1}-\eqref{eq:EinParamFits3} provide a good description of the final remnant, as a function of the ICs and the orbital parameters (specifically $\kappa$). 

%%%%%%FIGURE 15%%%%%%

%%%%%%%%%%%%%%%%%%%%%%%%%%%%%%%%%%%%%%%%%%%%%%%%%%
\section{Implications for Concentration Changes} \label{sec:Conc}

Having described how the profile  of the remnant depends on the ICs and the merger parameters, we will now consider the implications for the evolution of the concentration parameter. 

\subsection{Definitions of concentration} \label{sec:ConcMeas}

There are several methods for measuring concentration in simulated haloes; most commonly concentration is defined as $c=r_{\rm vir}/r_{-2}$, where $r_{-2}$ is determined through profile fitting and $r_{\rm vir}$ is defined in terms of the enclosed mean density. However, concentration can also be determined indirectly by measuring other properties, such as the peak circular velocity  \citep{prada2012, klypin2016} or the mean density profile \citep{alam2002, diemand2007}, and inverting the relationship between these properties and the usual concentration parameter, given an assumed theoretical profile. These different methods should all agree for NFW profiles, but can give different answers when the density profile is not NFW.

As discussed in \cite{prada2012} and \cite{klypin2016}, determining the concentration from the ratio of the peak circular velocity to the circular velocity at the virial radius,
\begin{equation} \label{eq:R}
R = v_{\rm peak}/v_{\rm vir} \,\,\,,
\end{equation}
may be more justified than profile fitting, since it accounts for differences in the Einasto shape parameter, $\alpha_{\rm E}$, that affect the central concentration of the mass distribution. Additionally, calculating $R$ does not require any assumption about the profile. 

Since $R$ is monotonically related to $c=r_{\rm vir}/r_{-2}$, to convert from $R$ to the concentration, $c_{\rm R}$, one can assume a profile (e.g. NFW or Einasto), and use the formula:
\begin{equation} \label{eq:cv_to_c}
R = \sqrt{\dfrac{c_{\rm R} f(x_{\rm peak})}{x_{\rm peak}f(c_{\rm R})}} \,\,\, ,
\end{equation}
where $f(x)$ is the mass profile corresponding to the specified density profile and $x_{\rm peak} \equiv r_{\rm peak}/r_{-2}$ \citep{klypin2016}. For convenience, we show how $R$ and $c_R$ are related in Appendix~\ref{sec:cM}.

 None of the usual methods for calculating $c$ are well defined in non-cosmological simulations, however, as they use a virial radius defined in terms of a mean background density. One approach to measuring relative {\it changes} in concentration, without needing to define a virial radius, is to scale the profiles so that they have the same $v_{\rm peak}$, and then scale $r_{\rm peak}$ by the same factor \citep{moore2004}. The concentration change can then be inferred from ratio of the intial $r_{\rm peak}$ to the shifted remnant $r_{\rm peak}'$. Mathematically, this can be expressed as:
\begin{equation}  \label{eq:cM}
\dfrac{c_M'}{c_M} =\dfrac{r_{\rm peak}}{r_{\rm peak}'} \dfrac{v_{\rm peak}'}{ v_{\rm peak}} \,\,\,,
\end{equation}
where $c_M'$ is the concentration of the remnant, and $c_M$ is the concentration of the original halo. For self-similar haloes with the same concentration, scaling the radii of the haloes so they have the same peak circular velocity ensures they have the same virial radius (since it accounts for the increase in mass). The radius at the peak circular velocity, $r_{\rm peak}$ then scales as $r_{-2}$ if the profile remains the same. The disadvantage to this method is that it assumes the remnant is self-similar to the initial profile, and that the background density remains constant. Since this definition is strongly dependent on the assumption of self-similar evolution, and we have demonstrated major mergers are not, in general, self-similar to the ICs, we will not use this definition to calculate concentration. However, for completeness, we present results using this definition in Appendix~\ref{sec:cM}.

\subsection{Concentration changes}

In this work, we track the net change in concentration from the ICs to the final remnant in three ways:
	\begin{enumerate}	
		\item using $c=r_{\rm vir}/r_{-2}$, with $r_{-2}$ measured directly from the logarithmic derivative of the density profile of the remnant, as the point where ${\rm d}\ln \rho/{\rm d}\ln r = -2$, smoothing with a Gaussian kernel of width $\sigma = 0.1$, while $r_{\rm vir}$ is defined based on the enclosed density, as explained in Section~\ref{subsec:structparam}.
		\item using $c_{\rm Ein}=r_{\rm vir}/r_{-2}$, with values of $r_{-2}$ and $r_{\rm vir}$  determined by fitting an Einasto profile to the individual merger remnant, as described in Sections~\ref{subsec:4.1}--\ref{subsec:4.2}.
		\item using $R=v_{\rm peak}/v_{\rm vir}$ with $v_{\rm peak}$ and $v_{\rm vir}$ determined directly from the density profile of the individual remnant, as in Section~\ref{subsec:structparam}.
	\end{enumerate}

Since the remnants are better described by Einasto profiles, we will not consider the concentration measured from the NFW fit, $c_{\rm NFW}$; however, we provide a comparison between $c_{\rm Ein}$ and $c_{\rm NFW}$ in  Appendix~\ref{sec:cM}.

In Fig.~\ref{fig:c_vs_Energy}, we show how the concentration parameters $c$ (method i), $c_{\rm Ein}$ (method ii), or $R$ (method iii) change with energy. As with the changes in the Einasto parameters, there appears to be a dependence on the initial halo model, particularly for $c_{\rm Ein}$. Individual concentration measurements that are based on direct measurements of $r_{-2}$ show that concentration changes decrease with $\kappa$; high-energy (low-$\kappa$) mergers cause an \emph{increase} in concentration, while low-energy  (high-$\kappa$) mergers decrease concentration. The trends in $c'_{\rm Ein}/c_{\rm Ein}$ and $R'/R$ are more similar to the expected result, with little change in concentration at low-$\kappa$, and increasing at high-$\kappa$ encounters . The ratio $c'/c$ obtained from direct measurements of the scale radius does not show the same increase in concentration at high-$\kappa$ values as the ratio derived from the Einasto fits; this could be because the haloes produced by mergers of very bound pairs are not as well approximated by the Einasto profile. From this plot, however, it appears that the profile-independent measurement of $R'/R$ shows a similar increase in concentration for high-$\kappa$ simulations, so it is more likely due to the numerical difficulty with measuring $c'/c$ directly. Finally, we note that $R$ can be mapped on to an equivalent (radial) concentration parameter $c_{\rm R}$; testing this inversion, we find that while $c_{\rm Ein}$ and $c_{\rm R}$ agree roughly, there is considerable ($\sim$20\%) scatter between the two. 

%The parameters $c$ and $c_{\rm Ein}$ can decrease for the remnants of low-energy mergers (large $\kappa$ values), especially those with the very extended EinLow profile, but generally increase by 10--20\%\ in most of our simulations. $R$ remains almost constant; the only exceptions are the very high-$\kappa$ simulations noted previously in Section~\ref{subsec:4.2}. As with the changes in the Einasto parameters, there appears to be a dependence on the initial halo model, particularly for $c_{\rm Ein}$. The trends in $c'_{\rm Ein}/c_{\rm Ein}$ and $R'/R$ look somewhat similar. The ratio $c'/c$ obtained from direct measurements of the scale radius (method (i)) does not show the same increase in concentration at high $\kappa$ values as the ratio derived from the Einasto fits (method (ii)); this could be because the haloes produced by mergers of very bound pairs are not as well approximated by the Einasto profile. From this plot, however, it appears that the profile-independent measurement of $R'/R$ shows a similar increase in concentration for high $\kappa$ simulations, so it is more likely due to the numerical difficulty with measuring $c'/c$ directly. Finally, we note that $R$ can be mapped on to an equivalent (radial) concentration parameter $c_{\rm R}$; testing this inversion, we find that while $c_{\rm Ein}$ and $c_{\rm R}$ agree roughly, there is considerable ($\sim$20\%) scatter between the two. 

%%%%%%FIGURE 18%%%%%%
\begin{figure*}
	\includegraphics[trim={0 0 0 0},clip]{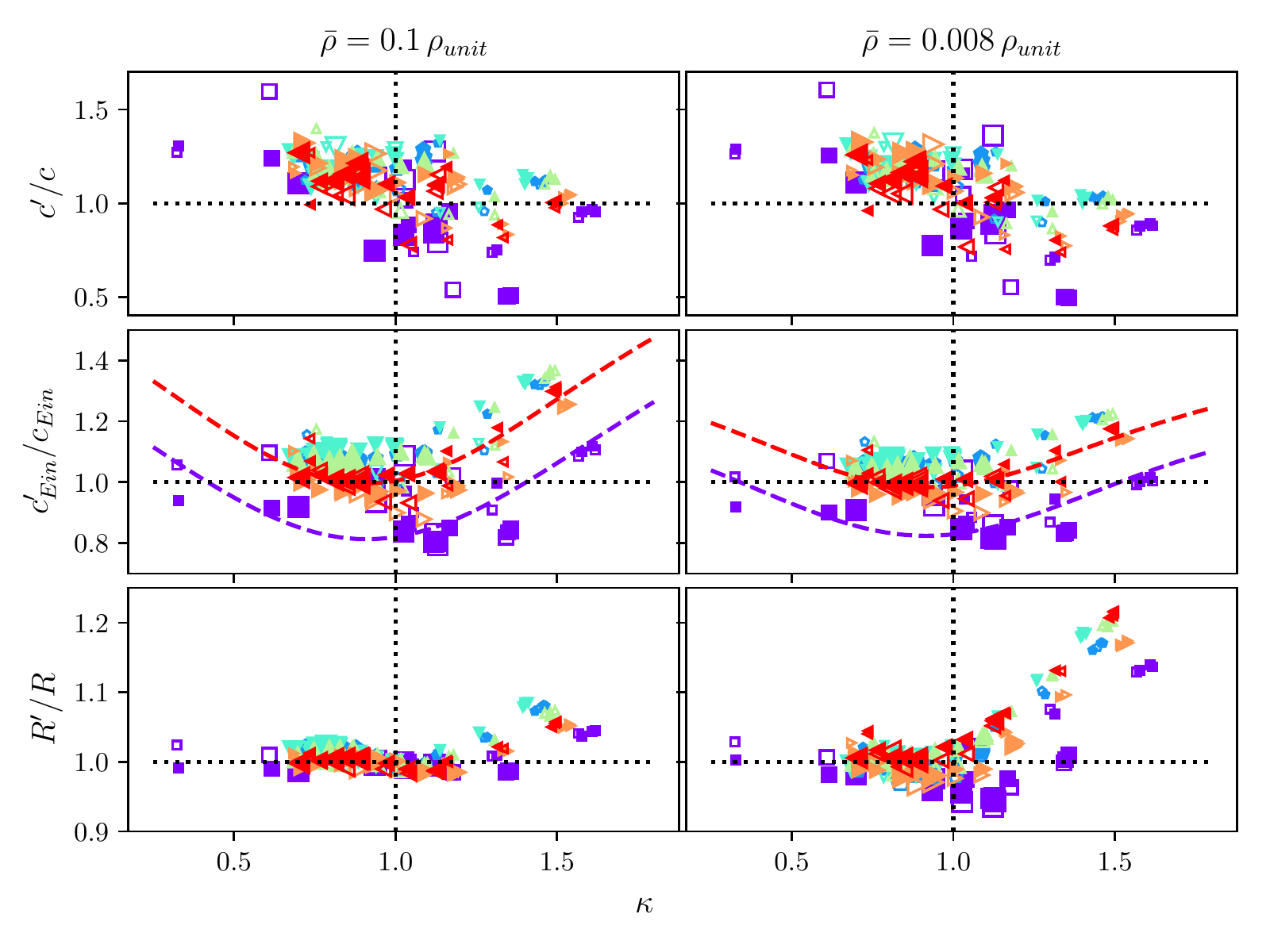}.
	\caption{Net concentration change as a function of the relative energy change, $\kappa$. The concentrations are determined using either scale radii measured directly for individual profiles (top row), from the Einasto parameters determined from profile fitting  (middle row), or derived from the velocity ratio $R=v_{\rm peak}/v_{\rm vir}$ measured directly from the profile (bottom row). The virial radii were calculated using overdensities $\bar{\rho} = 0.1\,\rho_{\rm unit}$ (left) and $0.008\,\rho_{\rm unit}$ (right). The dashed curves indicate $c'=c_{\rm analytic}$ recovered from the fits given in the text  (equations~\ref{eq:EinParamFits1}--\ref{eq:EinParamFits3}), for two values of $E_0$, as described in the following section. Colours and symbols are as in Fig.~\ref{fig:OrbitalParameters2}.}
	\label{fig:c_vs_Energy}
\end{figure*}

Overall, the pattern of concentration changes is complicated, depending on both $\kappa$ and the initial halo model. Concentration measurements are very sensitive to the method used, and there is considerable scatter from one method to another. Regardless of the method used, however, we find that concentration usually {\it increases} in major mergers, except in the case of mergers between very bound pairs and/or between haloes with low-$\alpha_{\rm E}$ parameters. It is surprising to find that high-energy major mergers rarely decrease concentration. This is somewhat contrary to the results by \cite{moore2004}, who found that two NFW haloes merging on a high-energy orbit with a tangential velocity produced a remnant that was less concentrated than its progenitors. On the other hand, it is more consistent with the results of \cite{kazantzidis2006}, as discussed further in Section~\ref{sec:Discuss}.

\subsection{Analytic model for concentration changes}

Ultimately, we wish to predict how concentration will change in major mergers. Our results indicate that the remnants of binary mergers are not, in general, self-similar to the ICs, but have properties that vary systematically with the energy of the system. Rather than determine concentration changes from direct fits to $r_{-2}$, together with some arbitrary choice of virial radius, we will use the analytic model from Section~\ref{subsec:4.3} to estimate how the profile will change, and then use that to calculate the corresponding change in concentration.

Given the initial haloes, and the orbital parameters of the merger, a prediction for the resulting Einasto remnant can be determined from the analytic predictions for the Einasto parameters. Then, the virial radius can be calculated from solving $\Delta_{\rm vir} \rho_b = \bar{\rho}(r_{\rm vir}) $, where $\bar{\rho}$ is the mean enclosed density of the Einasto profile. The radius of the peak circular velocity can be found from the approximation $r_{\rm peak} \approx 3.15 \exp{(-0.64 \alpha_{\rm E}^{1/3})} r_{-2}$ \citep{klypin2016}, and $v_{\rm peak}$ and $v_{\rm vir}$ from the circular velocity profile $v_{\rm circ}=\sqrt{GM/r}$. In what follows, we refer to concentrations predicted by this analytic model as $c_{\rm analytic}  = r_{\rm vir}/r_{-2}$ or $R_{\rm analytic} = v_{\rm peak}/v_{\rm vir}$.

\begin{figure*}
	\includegraphics[trim={0 0 0 0},clip]{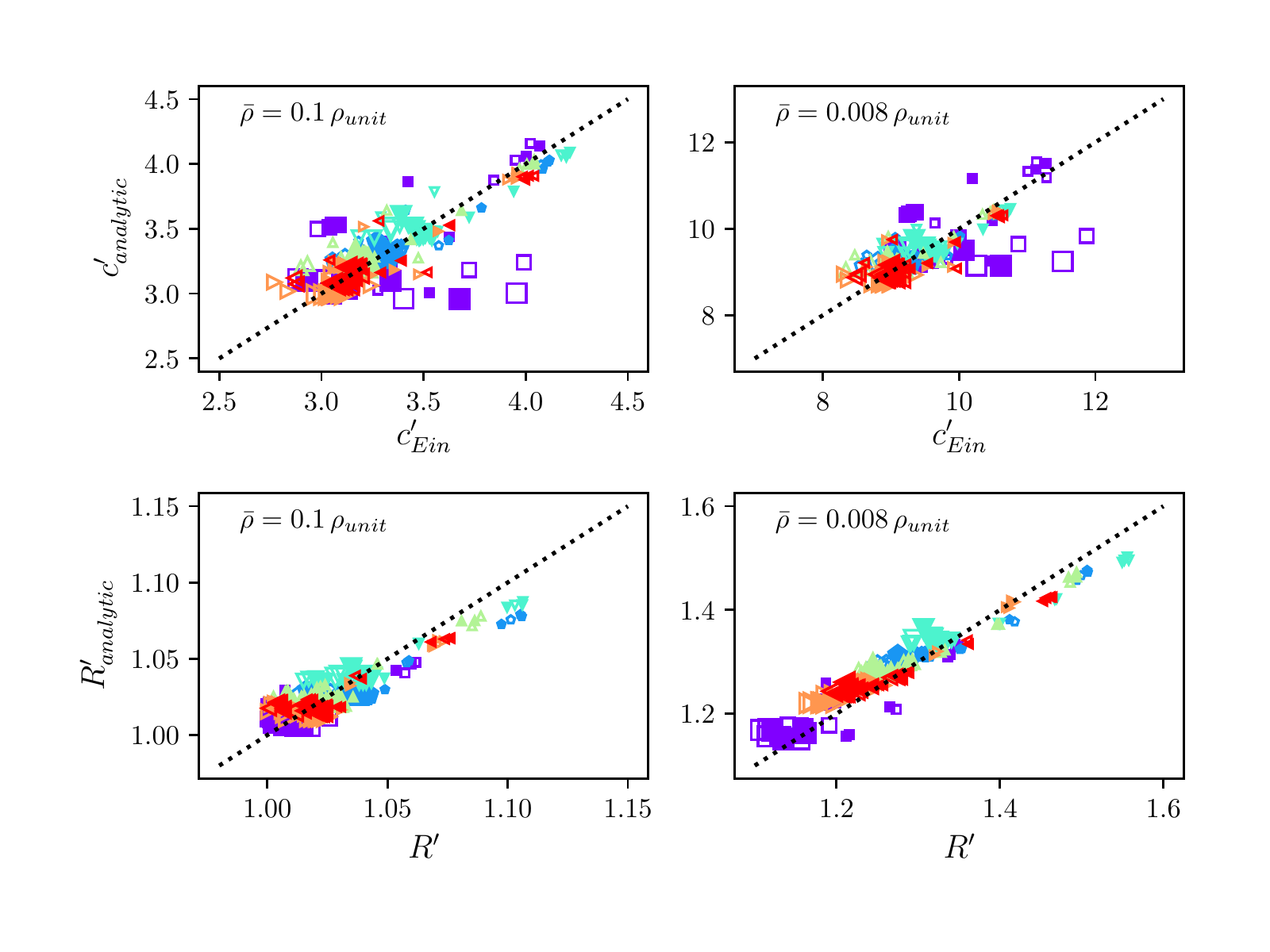}.
	\caption{Predictions for concentration parameter (top) and velocity ratio $R$ (bottom) derived from our analytic fit (equations~\ref{eq:EinParamFits1}--\ref{eq:EinParamFits3}), versus values measured from individual profiles. The virial radii were calculated using overdensities $\bar{\rho} = 0.1\,\rho_{\rm unit}$ (left) and $0.008\,\rho_{\rm unit}$ (right). Colours and symbols are as in Fig.~\ref{fig:OrbitalParameters2}.}
	\label{fig:ModelFits_conc}
\end{figure*}

Fig.~\ref{fig:ModelFits_conc} compares concentration values derived from the predicted profile changes using equations~\eqref{eq:EinParamFits1}-\eqref{eq:EinParamFits3} to concentrations measured from Einasto fits to individual profiles, for two choices of the virial radius. On the whole the two values agree fairly well; the RMS scatter between results and fit are 0.2 (top left), 0.5 (top right), 0.01 (bottom left), and 0.03 (bottom right), although once again much of this comes from the EinLow (square) simulations. From this plot it appears that the $R'$ prediction, $R'_{\rm analytic}$, is more successful than the $c'$ prediction, $c'_{\rm analytic}$. However, once the $R$ values are mapped back to $c$ (a mapping that is sensitive to the predicted profile), this is no longer the case. Overall, the concentration predictions presented here are accurate to within approximately 10\%.

%%%%%%FIGURE 19%%%%%%

\subsection{Implications for the boost factor}

Concentration changes have important consequences for the central densities of haloes, and therefore the dark matter annihilation boost factor. The boost factor within a volume $V$ is defined as:
\begin{equation} \label{eq:boost}
B = \dfrac{1}{\bar{\rho}^2 V} \int \rho^2 {\rm d}V \,\,\, ,
\end{equation}
where $\rho$ is the density of the halo;  \citep[see, e.g.,][for a discussion]{okoli2018}.

\begin{figure}
	\includegraphics[width=\columnwidth,trim={3.5cm 0cm 5cm 0cm},clip]{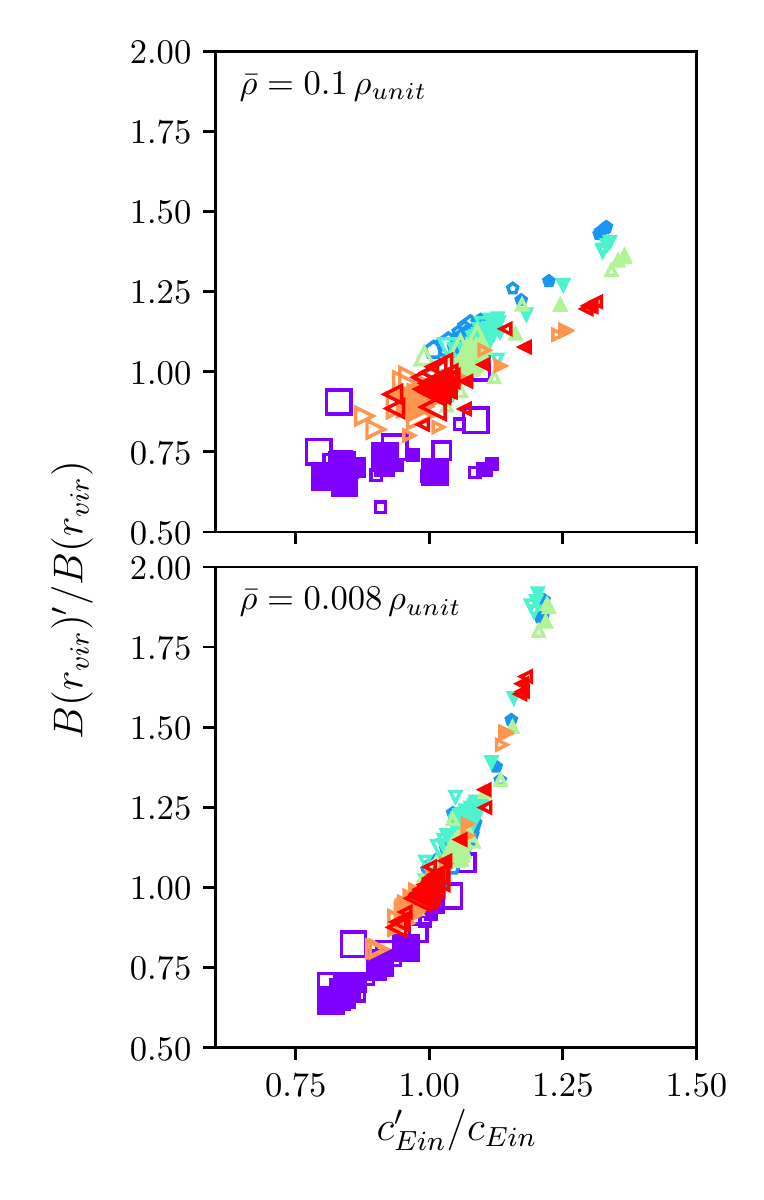}.
	\caption{ Relation between the change in boost factor and the change in concentration calculated as $c=r_{\rm vir}/r_{-2}$ from the best Einasto fit. The virial radii were calculated using overdensities $\bar{\rho} = 0.1\,\rho_{\rm unit}$ (top) and $0.008\,\rho_{\rm unit}$ (bottom). Colours and symbols are as in Fig.~\ref{fig:OrbitalParameters2}.}
	\label{fig:Boost_vs_c}
\end{figure}

Since this calculation can be sensitive to the inner regions of the profile, we calculate the boost factor within the virial radius from the best-fitting Einasto profile, assuming spherical symmetry in the remnant. Fig.~\ref{fig:Boost_vs_c} shows that the change in boost factor is correlated with the change in concentration, as expected. We have compared these direct calculations of the boost factor to the values obtained assuming our scaling relations for the Einasto parameters, and find that they agree to within approximately 5\%\ .

%%%%%%FIGURE 16%%%%%%

%%%%%%%%%%%%%%%%%%%%%%%%%%%%%%%%%%%%%%%%%%%%%%%%%%%%%%%%%%%%%%%%%
\section{Discussion} \label{sec:Discuss}

A number of lines of evidence suggest that major mergers play an important role in determining the properties of dark matter haloes. In this paper, we have performed over a hundred simulations of major mergers between identical, isolated haloes with various density profiles and initial orbits, and have tracked how the density profile of the remnant differs from that of the initial haloes. The differences are subtle, but indicate that the evolution from the initial to the final state is not self-similar, although the remnants are well described by Einasto profiles. Relative to scaled ICs, the mass distributions of the haloes are rearranged in a systematic way, with low-energy (high-$\kappa$) mergers resulting in the mass moving inwards to higher density, while high-energy (low-$\kappa$) mergers result in more extended haloes. Some halo properties depend mainly on $\kappa$ (particularly the half-mass radius and the peak circular velocity), while others also depend on the initial halo model (e.g. the Einasto parameters $\alpha_{\rm E}$, $r_{-2}$, and $\rho_{-2}$). 

A surprising result of our study is that while energetic mergers produce more extended mass distributions, they do not generally reduce the concentration parameter significantly,  and they result in an increased central density, even compared to the expectation from self-similar scaling. In addition, although the scale radius generally increases after mergers, it does not increase as much as expected from self-similar evolution (e.g. Fig.~\ref{fig:Einparams}). This suggests that major mergers do not fully explain the evolution of the central density or scale radius of haloes as they grow.

Considering the evolution of the halo mass profiles in detail, we find significant rearrangement of the mass distribution. If we define a virial radius corresponding to the density contour $\bar{\rho}=0.008\, \rho_{\rm unit}$, for instance, we find that the final virial mass ranges from 1.3 to 4.1 times that of the ICs, where the expected value for self-similar evolution is 2. Previous studies of isolated major mergers have found that about 20--50 per cent of the mass lies outside the virial radius \citep[e.g.][]{kazantzidis2006, valluri2007, vass2009}. Our results are consistent with these previous ones, but we have tested a much larger range of orbital parameters, including unrealistically low-energy mergers in which the virial mass increases more than expected from self-similar evolution. The fact that virial mass is not additive in halo mergers has important implications for semi-analytic models of galaxy formation, as discussed in \cite{kazantzidis2006} and \cite{valluri2007}. Similarly, as pointed out in \cite{okoli2018}, boost factor calculations often assume that the virial mass is additive; relaxing this assumption would affect boost factor predictions considerably.

%These findings also have implications for the study of the origin of the universal density profile. Our results are inconsistent with the idea that the logarithmic inner slope of halo density profiles should tend towards $-1$, as expected if the NFW model is a dynamical attractor, but instead agree with phase-space arguments  \citep{dehnen2005} and recent simulation results \citep[e.g.][]{lebrun2018} that suggest that the steepest cusp is preserved in mergers. 

One of our original goals was to produce a prediction for how halo concentration changes in major mergers. Using fits to the evolution of the three Einasto parameters shown in equations~\eqref{eq:EinParamFits1}-\eqref{eq:EinParamFits3}, we have produced predictions that match our simulation results to 10\%\ on average. 
%In most of our simulations, the scale radius $r_{-2}$ increases {\it less} than expected for self-similar evolution. Self-similar evolution corresponds to a halo keeping the same concentration as it increases in mass; thus in most of our simulations the concentration parameter {\it increases}. 
Overall, it seems that while major mergers can cause systematic departures from self-similar evolution and corresponding changes in concentration, the pattern is opposite to what one might naively expect, in that energetic mergers (low-$\kappa$ values) {\it increase the concentration} while low-energy (tightly bound) mergers decrease it. This is in contrast to \cite{moore2004}, who concluded (from a single example) that high-energy mergers resulting in oblate haloes may decrease halo concentration. This could be because they assumed self-similar evolution in their concentration measurement. Our results are more consistent with \cite{kazantzidis2006}, who concluded that haloes are robust to major mergers. Though they also found that the profile changes  resulting from major mergers are subtle, their fig.~8 suggests that mass moves inwards for haloes with lower central densities, but outwards for haloes with high central density compared to the scaled initial models. This is similar to our results, where the high-central density EinLow haloes tended to produce remnants with decreased concentrations.

Given our results, if the central density in haloes does drop as they grow, the mechanism for this remains unclear. This could be because we have only considered the simplest model for major mergers; binary, equal-mass mergers of identical haloes that are spherical, non-rotating, and lack substructure. How these results extend to more realistic mergers in a cosmological context is also unclear; cosmological conditions are different because equal-mass and/or isolated binary mergers are rare. Given the merger remnant takes an orbital period or two to settle into virial equilibrium, multiple mergers staggered by some fraction of an orbital period might lead to a decrease in central density, or substructure left by an earlier merger may affect the evolution of a later one. We will consider the effect of multiple mergers in future work.

In conclusion, we have explored how the density profiles and concentrations of dark matter haloes change in equal-mass mergers. These mergers do not generally decrease the central density or the concentration parameter, and often cause remnants to become more concentrated. This is a puzzle, given previous results suggesting the central density must drop as haloes grow. Therefore mergers seem unlikely to cause the evolution of halo central density needed to explain results from cosmological simulations, as pointed out by \cite{okoli2018}. Our future work will explore more complicated and more realistic merger scenarios, to see if the trends found in this paper still hold in a cosmological setting. 

\section*{Acknowledgements}
NED acknowledges support from NSERC Canada, through a postgraduate scholarship. JET acknowledges financial support from NSERC Canada, through a Discovery Grant. The authors also wish to thank Michael Hudson, Julio Navarro, and the anonymous referee for useful comments.

\appendix

\section{Inner Slope}\label{sec:innerslope}

\begin{figure}
	\includegraphics[width=\columnwidth,trim={0cm 5.5cm 8cm 0},]{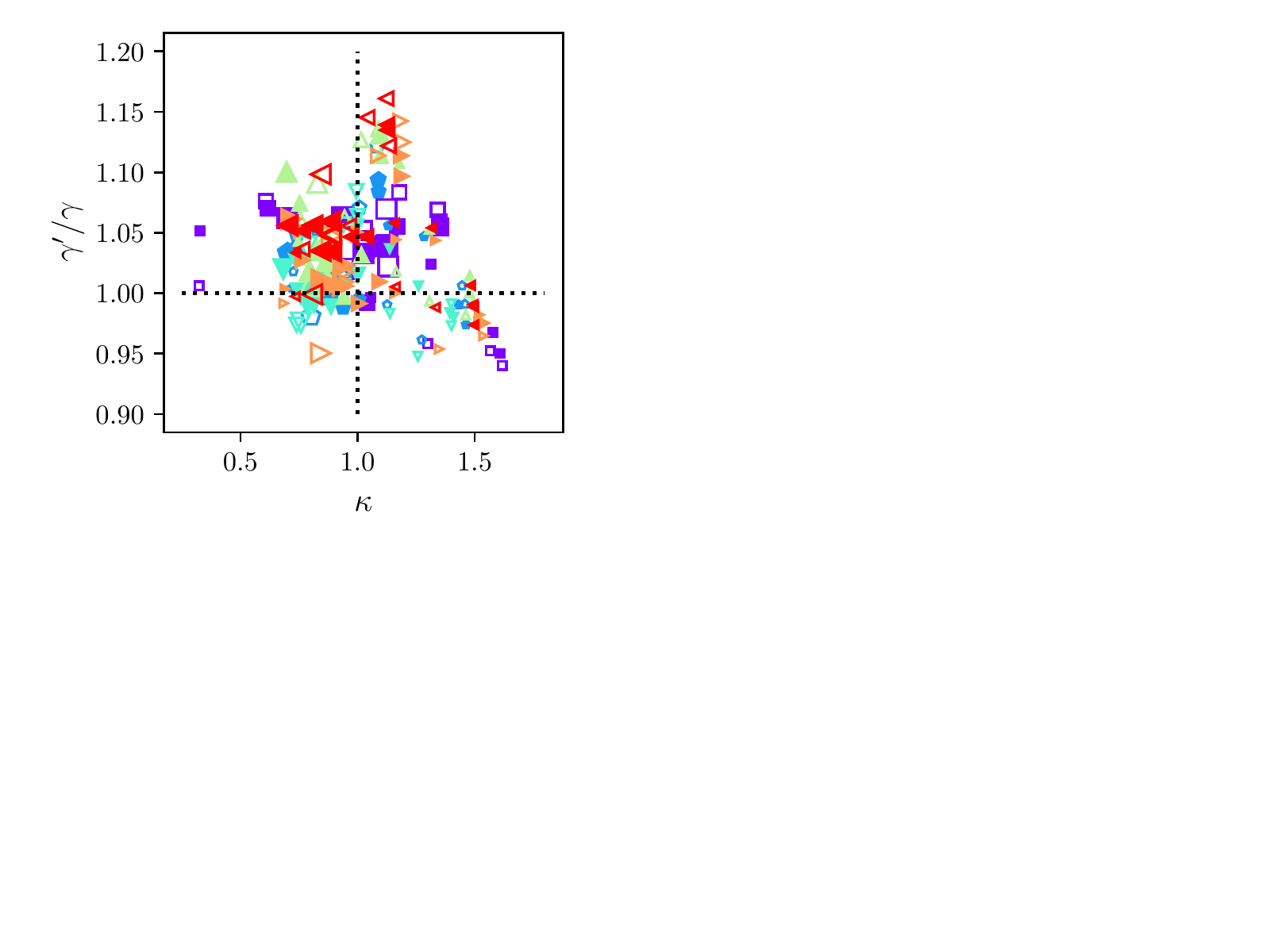}.
	\caption{The change in the inner slope, $\gamma'/\gamma$, calculated between $r = 0.1\,r_{\rm peak}$ and $r = 0.4\,r_{\rm peak}$ as a function of the relative energy change $\kappa$. The dotted lines are the expectations for self-similar evolution of the density profile. Colours and symbols are as in Fig.~\ref{fig:OrbitalParameters2}.}
	\label{fig:InnerSlope}
\end{figure}

Fig.~\ref{fig:InnerSlope} shows changes in the inner slope, $\gamma'/\gamma$, calculated between $r = 0.1\,r_{\rm peak}$ and $r = 0.4\,r_{\rm peak}$. This quantity does not seem to scale in a predictable way with the orbital parameters or the change in internal energy. There is potentially a trend with the initial separation of the haloes, $r_{\rm sep}$ (represented by the size of the symbols). This is consistent with the findings from \cite{ogiya2016}, who found that the change in inner slope depends on $r_{\rm sep}$. 

%%%%%%FIGURE A1%%%%%%

\begin{figure}
	\includegraphics[width=\columnwidth,trim={0cm 5cm 8cm 0},clip]{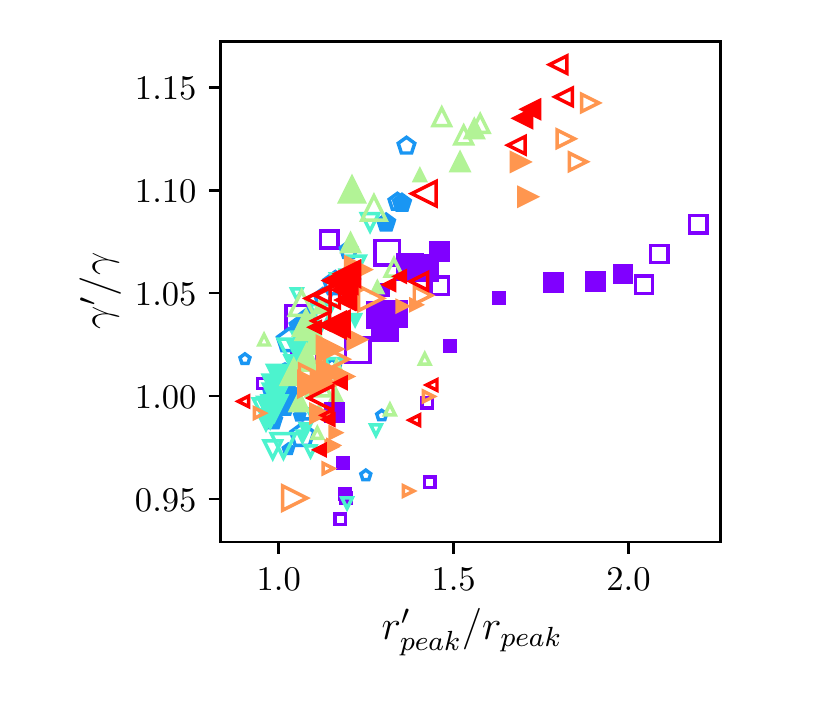}.
	\caption{The change in the inner slope, $\gamma'/\gamma$, calculated between $r = 0.1\,r_{\rm peak}$ and $r = 0.4\,r_{\rm peak}$ compared to the change in the radius corresponding to the peak in the circular velocity curve $r_{\rm peak}'/r_{\rm peak}$. Colours and symbols are as in Fig.~\ref{fig:OrbitalParameters2}.}
	\label{fig:InnerSlope_vs_rpeak}
\end{figure}

The relation between $\gamma'/\gamma$ and $\kappa$ is somewhat similar to that between $r_{\rm peak}'/r_{\rm peak}$ and $\kappa$. Therefore, we also show the relation between $\gamma'/\gamma$ and $r_{\rm peak}'/r_{\rm peak}$ in Fig.~\ref{fig:InnerSlope_vs_rpeak}. This demonstrates a correlation between these two parameters, suggesting that the complicated changes in $r_{\rm peak}$ may be due to a combination of mass rearrangement and evolution of the inner slope of the remnant.
%%%%%%FIGURE A2%%%%%%

\section{Comparison of Different Concentration Measurements} \label{sec:cM}

\begin{figure}
	\includegraphics[trim={0.5cm 0cm 7.5cm 0cm},clip,width=\columnwidth]{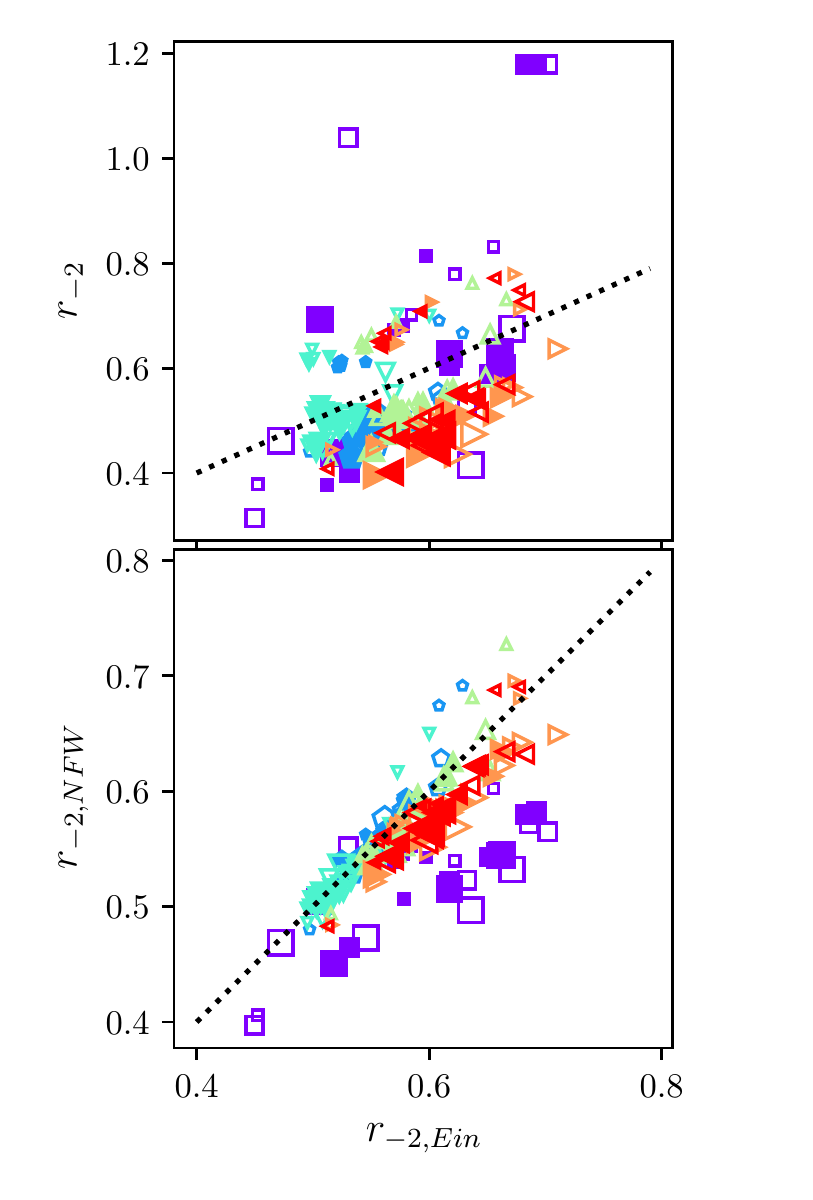}.
	\caption{Comparison of the scale radius measured directly from the slope of the density profile (top), and determined from the NFW fit (bottom), 
		to the value determined from the Einasto fit. The dotted lines indicate a 1--1 relation. Colours and symbols are as in Fig.~\ref{fig:OrbitalParameters2}.}
	\label{fig:r2_compare}
\end{figure}

As the traditional definition of the concentration parameter depends on the scale radius $r_{-2}$, in Fig.~\ref{fig:r2_compare} we compare the values measured directly from the density profile, as explained in Section~\ref{subsec:structparam}, to the values derived from the NFW and Einasto fits. The Einasto fit predicts larger scale radii than the direct measurement, except in some of the simulations with low initial radial separations, and in general there is considerable scatter between the two sets of measurements. The  values of $r_{-2}$ from Einasto and NFW profile fits are in much better agreement, except for the case of the EinLow simulations, where the Einasto fit predicts systematically larger scale radii.

%%%%%%FIGURE 12%%%%%%

\begin{figure}
	\includegraphics[width=\columnwidth,trim={3.5cm 0cm 4.5cm 0cm},clip]{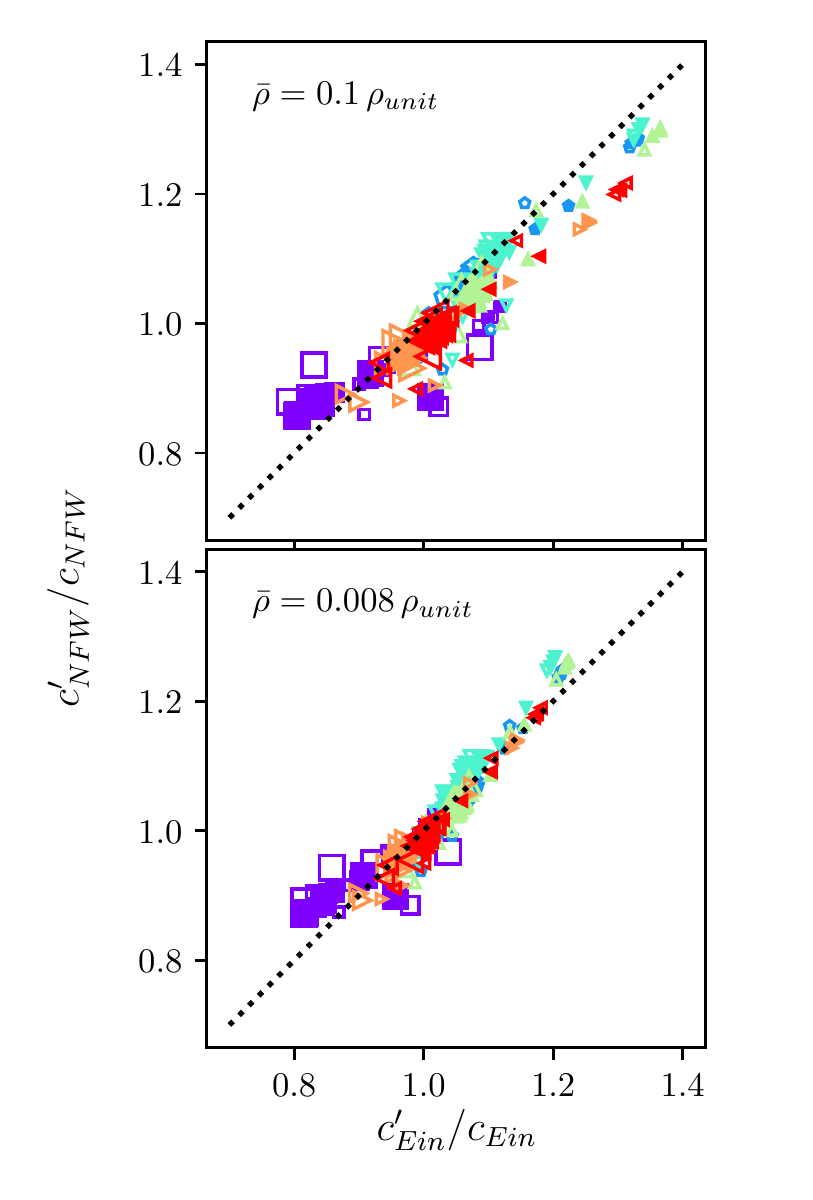}.
	\caption{A comparison of the concentration change calculated from the NFW fit, $c = r_{\rm vir}/r_{s}$ versus the concentrations calculated from the Einasto fit, $c = r_{\rm vir}/r_{-2}$. The virial radii were calculated using overdensities $\bar{\rho} = 0.1\,\rho_{\rm unit}$ (top) and $0.008\,\rho_{\rm unit}$ (bottom). The dotted line shows where the two definitions are equal. Colours and symbols are as in Fig.~\ref{fig:OrbitalParameters2}.}
	\label{fig:cEin_vs_cNFW}
\end{figure}

Additionally, we compare concentration measured through either NFW or Einasto fits in Fig.~\ref{fig:cEin_vs_cNFW}. The two methods roughly agree, though there is some scatter. Some of this scatter may be from systematic errors in NFW fits; it has been shown that NFW fits overpredict the halo concentration by 10--20\% for high-$\nu$ haloes \citep{klypin2016}. We find a similar result; although changes in concentration measured from NFW fits are slightly lower than from Einasto fits, the actual concentration values are higher for the NFW fits, particularly for the EinLow simulations.

%%%%%%FIGURE 17%%%%%%

\begin{figure}
	\includegraphics[width=\columnwidth,trim={3.5cm 0 4.5cm 0},clip]{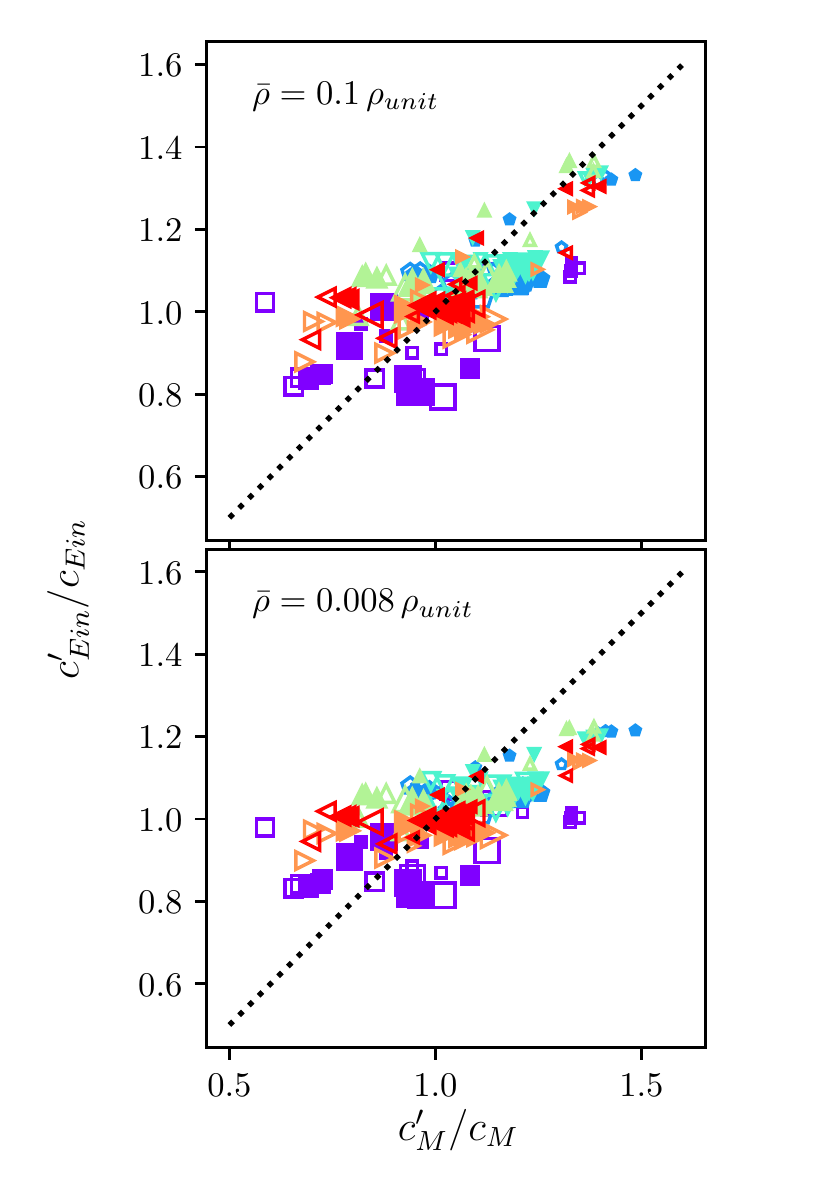}.
	\caption{A comparison of the concentration change calculated from the Einasto fit, $c_{\rm Ein}=r_{\rm vir}/r_{-2}$ versus the concentration calculated as $c_M$ (see equation~\eqref{eq:cM}). The dotted line shows where the two definitions are equal. Top and bottom panels use the overdensity of an NFW profile of concentration 3 ($\bar{\rho}=0.1\,\rho_{\rm unit}$) and 10  ($\bar{\rho}=0.008\,\rho_{\rm unit}$)  to calculate the virial radii, respectively. Colours and symbols are as in Fig.~\ref{fig:OrbitalParameters2}.}
	\label{fig:c_vs_cM}
\end{figure}

\begin{figure}
	\includegraphics[width=\columnwidth,trim={3.5 5.5cm 7.5cm 0},clip]{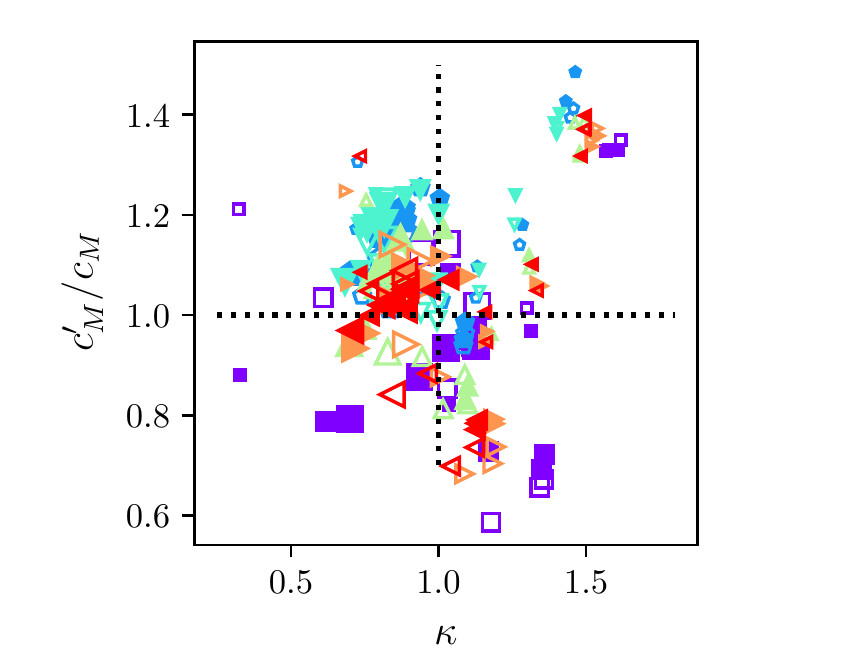}.
	\caption{Concentration change $c_M'/c_M$ as a function of the relative energy change, $\kappa$. Colours and symbols are as in Fig.~\ref{fig:OrbitalParameters2}.}
	\label{fig:cM_vs_Energy}
\end{figure}

Since the halo remnants presented in this study are not, in general, self-similar to the ICs, we do not expect the relative change in concentration calculated using the method of \cite{moore2004}, $c_M'/c_M$, to match the actual value found by profile fitting. To demonstrate this, we compare  $c_M'/c_M$ to the change in concentration measured through $c_{\rm Ein}=r_{\rm vir}/r_{-2}$ 
(calculated from the best Einasto fit) in Fig.~\ref{fig:c_vs_cM}. As expected, there is a large discrepancy between these two measurements, further emphasizing that the profile remnants do not evolve in a self-similar manner. In Fig.~\ref{fig:cM_vs_Energy}, we also show $c_M'/c_M$ as a 
function of $\kappa$.

%%%%%%FIGURE A3%%%%%%

%%%%%%FIGURE A4%%%%%%

\begin{figure}
	\includegraphics[width=\columnwidth,scale = 0.7,trim={0.5cm 5cm 8cm 1cm},clip]{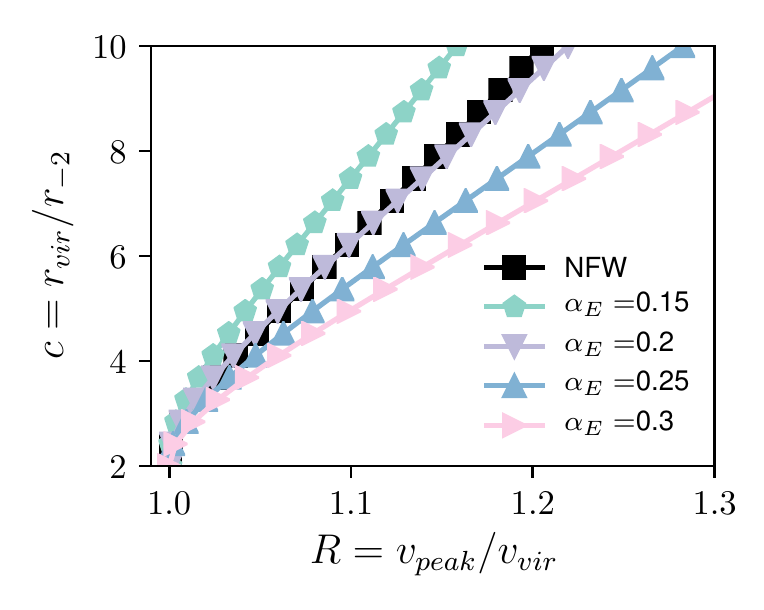}.
	\caption{Relationship between $R=v_{\rm peak}/v_{\rm vir}$ and the concentration, $c=r_{\rm vir}/r_{-2}$, for an NFW profile and Einasto profiles with varying $\alpha_{\rm E}$ parameters.}
	\label{fig:c_vs_R}
\end{figure}

Finally, in Fig.~\ref{fig:c_vs_R} we show the concentration, $c$, as a function of $R$ for NFW and various Einasto profiles. To calculate $f(x_{\rm peak})$, for Einasto profiles, we used the approximation $x_{\rm peak} \approx 3.15 \exp{(-0.64 \alpha_{\rm E}^{1/3})}$ \citep{klypin2016}.

%%%%%%FIGURE 16%%%%%%

\clearpage
\bibliographystyle{mnras}
\bibliography{MajorMerger_Conc}

\begin{thebibliography}{}
\makeatletter
\relax
\def\mn@urlcharsother{\let\do\@makeother \do\$\do\&\do\#\do\^\do\_\do\%\do\~}
\def\mn@doi{\begingroup\mn@urlcharsother \@ifnextchar [ {\mn@doi@}
  {\mn@doi@[]}}
\def\mn@doi@[#1]#2{\def\@tempa{#1}\ifx\@tempa\@empty \href
  {http://dx.doi.org/#2} {doi:#2}\else \href {http://dx.doi.org/#2} {#1}\fi
  \endgroup}
\def\mn@eprint#1#2{\mn@eprint@#1:#2::\@nil}
\def\mn@eprint@arXiv#1{\href {http://arxiv.org/abs/#1} {{\tt arXiv:#1}}}
\def\mn@eprint@dblp#1{\href {http://dblp.uni-trier.de/rec/bibtex/#1.xml}
  {dblp:#1}}
\def\mn@eprint@#1:#2:#3:#4\@nil{\def\@tempa {#1}\def\@tempb {#2}\def\@tempc
  {#3}\ifx \@tempc \@empty \let \@tempc \@tempb \let \@tempb \@tempa \fi \ifx
  \@tempb \@empty \def\@tempb {arXiv}\fi \@ifundefined
  {mn@eprint@\@tempb}{\@tempb:\@tempc}{\expandafter \expandafter \csname
  mn@eprint@\@tempb\endcsname \expandafter{\@tempc}}}

\bibitem[\protect\citeauthoryear{{Aceves} \& {Vel{\'a}zquez}}{{Aceves} \&
  {Vel{\'a}zquez}}{2006}]{aceves2006}
{Aceves} H.,  {Vel{\'a}zquez} H.,  2006, \rmxaa, \href
  {http://adsabs.harvard.edu/abs/2006RMxAA..42...41A} {42, 41}

\bibitem[\protect\citeauthoryear{{Alam}, {Bullock}  \& {Weinberg}}{{Alam}
  et~al.}{2002}]{alam2002}
{Alam} S.~M.~K.,  {Bullock} J.~S.,   {Weinberg} D.~H.,  2002, \mn@doi [\apj]
  {10.1086/340190}, \href {http://adsabs.harvard.edu/abs/2002ApJ...572...34A}
  {572, 34}

\bibitem[\protect\citeauthoryear{{Binney} \& {Tremaine}}{{Binney} \&
  {Tremaine}}{1987}]{binney}
{Binney} J.,  {Tremaine} S.,  1987, {Galactic dynamics}

\bibitem[\protect\citeauthoryear{{Boylan-Kolchin} \& {Ma}}{{Boylan-Kolchin} \&
  {Ma}}{2004}]{boylankolchin2004}
{Boylan-Kolchin} M.,  {Ma} C.-P.,  2004, \mn@doi [\mnras]
  {10.1111/j.1365-2966.2004.07585.x}, \href
  {http://adsabs.harvard.edu/abs/2004MNRAS.349.1117B} {349, 1117}

\bibitem[\protect\citeauthoryear{{Bullock}, {Kolatt}, {Sigad}, {Somerville},
  {Kravtsov}, {Klypin}, {Primack}  \& {Dekel}}{{Bullock}
  et~al.}{2001}]{bullock2001}
{Bullock} J.~S.,  {Kolatt} T.~S.,  {Sigad} Y.,  {Somerville} R.~S.,  {Kravtsov}
  A.~V.,  {Klypin} A.~A.,  {Primack} J.~R.,   {Dekel} A.,  2001, \mn@doi
  [\mnras] {10.1046/j.1365-8711.2001.04068.x}, \href
  {http://adsabs.harvard.edu/abs/2001MNRAS.321..559B} {321, 559}

\bibitem[\protect\citeauthoryear{{Diemand}, {Kuhlen}  \& {Madau}}{{Diemand}
  et~al.}{2007}]{diemand2007}
{Diemand} J.,  {Kuhlen} M.,   {Madau} P.,  2007, \mn@doi [\apj]
  {10.1086/520573}, \href {http://adsabs.harvard.edu/abs/2007ApJ...667..859D}
  {667, 859}

\bibitem[\protect\citeauthoryear{{Diemer}, {More}  \& {Kravtsov}}{{Diemer}
  et~al.}{2013}]{diemer2013}
{Diemer} B.,  {More} S.,   {Kravtsov} A.~V.,  2013, \mn@doi [\apj]
  {10.1088/0004-637X/766/1/25}, \href
  {http://adsabs.harvard.edu/abs/2013ApJ...766...25D} {766, 25}

\bibitem[\protect\citeauthoryear{{Drakos}, {Taylor}  \& {Benson}}{{Drakos}
  et~al.}{2017}]{drakos2017}
{Drakos} N.~E.,  {Taylor} J.~E.,   {Benson} A.~J.,  2017, \mn@doi [\mnras]
  {10.1093/mnras/stx652}, \href
  {http://adsabs.harvard.edu/abs/2017MNRAS.468.2345D} {468, 2345}

\bibitem[\protect\citeauthoryear{{Drakos}, {Taylor}, {Berrouet}, {Robotham}  \&
  {Power}}{{Drakos} et~al.}{2018}]{drakos2018}
{Drakos} N.~E.,  {Taylor} J.~E.,  {Berrouet} A.,  {Robotham} A.~S.~G.,
  {Power} C.,  2018, preprint, \href
  {http://adsabs.harvard.edu/abs/2018arXiv181112839D} {} (\mn@eprint {arXiv}
  {1811.12839})

\bibitem[\protect\citeauthoryear{{Duffy}, {Schaye}, {Kay}  \& {Dalla
  Vecchia}}{{Duffy} et~al.}{2008}]{duffy2008}
{Duffy} A.~R.,  {Schaye} J.,  {Kay} S.~T.,   {Dalla Vecchia} C.,  2008, \mn@doi
  [\mnras] {10.1111/j.1745-3933.2008.00537.x}, \href
  {http://adsabs.harvard.edu/abs/2008MNRAS.390L..64D} {390, L64}

\bibitem[\protect\citeauthoryear{{Dutton} \& {Macci{\`o}}}{{Dutton} \&
  {Macci{\`o}}}{2014}]{dutton2014}
{Dutton} A.~A.,  {Macci{\`o}} A.~V.,  2014, \mn@doi [\mnras]
  {10.1093/mnras/stu742}, \href
  {http://adsabs.harvard.edu/abs/2014MNRAS.441.3359D} {441, 3359}

\bibitem[\protect\citeauthoryear{{Gao}, {Navarro}, {Cole}, {Frenk}, {White},
  {Springel}, {Jenkins}  \& {Neto}}{{Gao} et~al.}{2008}]{gao2008}
{Gao} L.,  {Navarro} J.~F.,  {Cole} S.,  {Frenk} C.~S.,  {White} S.~D.~M.,
  {Springel} V.,  {Jenkins} A.,   {Neto} A.~F.,  2008, \mn@doi [\mnras]
  {10.1111/j.1365-2966.2008.13277.x}, \href
  {http://adsabs.harvard.edu/abs/2008MNRAS.387..536G} {387, 536}

\bibitem[\protect\citeauthoryear{{Guo}, {Cole}, {Eke}  \& {Frenk}}{{Guo}
  et~al.}{2012}]{guo2012}
{Guo} Q.,  {Cole} S.,  {Eke} V.,   {Frenk} C.,  2012, \mn@doi [\mnras]
  {10.1111/j.1365-2966.2012.21882.x}, \href
  {http://adsabs.harvard.edu/abs/2012MNRAS.427..428G} {427, 428}

\bibitem[\protect\citeauthoryear{{Ishiyama}}{{Ishiyama}}{2014}]{ishiyama2014}
{Ishiyama} T.,  2014, \mn@doi [\apj] {10.1088/0004-637X/788/1/27}, \href
  {http://adsabs.harvard.edu/abs/2014ApJ...788...27I} {788, 27}

\bibitem[\protect\citeauthoryear{{Jing}}{{Jing}}{2000}]{jing2000}
{Jing} Y.~P.,  2000, \mn@doi [\apj] {10.1086/308809}, \href
  {http://adsabs.harvard.edu/abs/2000ApJ...535...30J} {535, 30}

\bibitem[\protect\citeauthoryear{{Kazantzidis}, {Zentner}  \&
  {Kravtsov}}{{Kazantzidis} et~al.}{2006}]{kazantzidis2006}
{Kazantzidis} S.,  {Zentner} A.~R.,   {Kravtsov} A.~V.,  2006, \mn@doi [\apj]
  {10.1086/500579}, \href {http://adsabs.harvard.edu/abs/2006ApJ...641..647K}
  {641, 647}

\bibitem[\protect\citeauthoryear{{Klypin}, {Yepes}, {Gottl{\"o}ber}, {Prada}
  \& {He{\ss}}}{{Klypin} et~al.}{2016}]{klypin2016}
{Klypin} A.,  {Yepes} G.,  {Gottl{\"o}ber} S.,  {Prada} F.,   {He{\ss}} S.,
  2016, \mn@doi [\mnras] {10.1093/mnras/stw248}, \href
  {http://adsabs.harvard.edu/abs/2016MNRAS.457.4340K} {457, 4340}

\bibitem[\protect\citeauthoryear{{McMillan}, {Athanassoula}  \&
  {Dehnen}}{{McMillan} et~al.}{2007}]{mcmillan2007}
{McMillan} P.~J.,  {Athanassoula} E.,   {Dehnen} W.,  2007, \mn@doi [\mnras]
  {10.1111/j.1365-2966.2007.11516.x}, \href
  {http://adsabs.harvard.edu/abs/2007MNRAS.376.1261M} {376, 1261}

\bibitem[\protect\citeauthoryear{{Meneghetti} et~al.,}{{Meneghetti}
  et~al.}{2014}]{meneghetti2014}
{Meneghetti} M.,  et~al., 2014, \mn@doi [\apj] {10.1088/0004-637X/797/1/34},
  \href {http://adsabs.harvard.edu/abs/2014ApJ...797...34M} {797, 34}

\bibitem[\protect\citeauthoryear{{Moore}, {Kazantzidis}, {Diemand}  \&
  {Stadel}}{{Moore} et~al.}{2004}]{moore2004}
{Moore} B.,  {Kazantzidis} S.,  {Diemand} J.,   {Stadel} J.,  2004, \mn@doi
  [\mnras] {10.1111/j.1365-2966.2004.08211.x}, \href
  {http://adsabs.harvard.edu/abs/2004MNRAS.354..522M} {354, 522}

\bibitem[\protect\citeauthoryear{{Navarro}, {Frenk}  \& {White}}{{Navarro}
  et~al.}{1996}]{navarro1996}
{Navarro} J.~F.,  {Frenk} C.~S.,   {White} S.~D.~M.,  1996, \mn@doi [\apj]
  {10.1086/177173}, \href {http://adsabs.harvard.edu/abs/1996ApJ...462..563N}
  {462, 563}

\bibitem[\protect\citeauthoryear{{Navarro}, {Frenk}  \& {White}}{{Navarro}
  et~al.}{1997}]{navarro1997}
{Navarro} J.~F.,  {Frenk} C.~S.,   {White} S.~D.~M.,  1997, \apj, \href
  {http://adsabs.harvard.edu/abs/1997ApJ...490..493N} {490, 493}

\bibitem[\protect\citeauthoryear{{Navarro} et~al.,}{{Navarro}
  et~al.}{2004}]{navarro2004}
{Navarro} J.~F.,  et~al., 2004, \mn@doi [\mnras]
  {10.1111/j.1365-2966.2004.07586.x}, \href
  {http://adsabs.harvard.edu/abs/2004MNRAS.349.1039N} {349, 1039}

\bibitem[\protect\citeauthoryear{{Neto} et~al.,}{{Neto}
  et~al.}{2007}]{neto2007}
{Neto} A.~F.,  et~al., 2007, \mn@doi [\mnras]
  {10.1111/j.1365-2966.2007.12381.x}, \href
  {http://adsabs.harvard.edu/abs/2007MNRAS.381.1450N} {381, 1450}

\bibitem[\protect\citeauthoryear{{Nusser} \& {Sheth}}{{Nusser} \&
  {Sheth}}{1999}]{nusser1999}
{Nusser} A.,  {Sheth} R.~K.,  1999, \mn@doi [\mnras]
  {10.1046/j.1365-8711.1999.02197.x}, \href
  {http://adsabs.harvard.edu/abs/1999MNRAS.303..685N} {303, 685}

\bibitem[\protect\citeauthoryear{{Ogiya}, {Nagai}  \& {Ishiyama}}{{Ogiya}
  et~al.}{2016}]{ogiya2016}
{Ogiya} G.,  {Nagai} D.,   {Ishiyama} T.,  2016, \mn@doi [\mnras]
  {10.1093/mnras/stw1551}, \href
  {http://adsabs.harvard.edu/abs/2016MNRAS.461.3385O} {461, 3385}

\bibitem[\protect\citeauthoryear{{Okoli}, {Taylor}  \& {Afshordi}}{{Okoli}
  et~al.}{2018}]{okoli2018}
{Okoli} C.,  {Taylor} J.~E.,   {Afshordi} N.,  2018, \mn@doi [\jcap]
  {10.1088/1475-7516/2018/08/019}, \href
  {http://adsabs.harvard.edu/abs/2018JCAP...08..019O} {8, 019}

\bibitem[\protect\citeauthoryear{{Ouellette} et~al.,}{{Ouellette}
  et~al.}{2017}]{ouellette2017}
{Ouellette} N.~N.-Q.,  et~al., 2017, \mn@doi [\apj] {10.3847/1538-4357/aa74b1},
  \href {http://adsabs.harvard.edu/abs/2017ApJ...843...74O} {843, 74}

\bibitem[\protect\citeauthoryear{{Peebles}}{{Peebles}}{1969}]{peebles1969}
{Peebles} P.~J.~E.,  1969, \mn@doi [\apj] {10.1086/149876}, \href
  {http://adsabs.harvard.edu/abs/1969ApJ...155..393P} {155, 393}

\bibitem[\protect\citeauthoryear{{Pilipenko}, {S{\'a}nchez-Conde}, {Prada}  \&
  {Yepes}}{{Pilipenko} et~al.}{2017}]{pilipenko2017}
{Pilipenko} S.~V.,  {S{\'a}nchez-Conde} M.~A.,  {Prada} F.,   {Yepes} G.,
  2017, \mn@doi [\mnras] {10.1093/mnras/stx2319}, \href
  {http://adsabs.harvard.edu/abs/2017MNRAS.472.4918P} {472, 4918}

\bibitem[\protect\citeauthoryear{{Prada}, {Klypin}, {Cuesta}, {Betancort-Rijo}
  \& {Primack}}{{Prada} et~al.}{2012}]{prada2012}
{Prada} F.,  {Klypin} A.~A.,  {Cuesta} A.~J.,  {Betancort-Rijo} J.~E.,
  {Primack} J.,  2012, \mn@doi [\mnras] {10.1111/j.1365-2966.2012.21007.x},
  \href {http://adsabs.harvard.edu/abs/2012MNRAS.423.3018P} {423, 3018}

\bibitem[\protect\citeauthoryear{{Springel}}{{Springel}}{2005}]{gadget2}
{Springel} V.,  2005, \mn@doi [\mnras] {10.1111/j.1365-2966.2005.09655.x},
  \href {http://adsabs.harvard.edu/abs/2005MNRAS.364.1105S} {364, 1105}

\bibitem[\protect\citeauthoryear{{Umetsu}, {Zitrin}, {Gruen}, {Merten},
  {Donahue}  \& {Postman}}{{Umetsu} et~al.}{2016}]{umetsu2016}
{Umetsu} K.,  {Zitrin} A.,  {Gruen} D.,  {Merten} J.,  {Donahue} M.,
  {Postman} M.,  2016, \mn@doi [\apj] {10.3847/0004-637X/821/2/116}, \href
  {http://adsabs.harvard.edu/abs/2016ApJ...821..116U} {821, 116}

\bibitem[\protect\citeauthoryear{{Valluri}, {Vass}, {Kazantzidis}, {Kravtsov}
  \& {Bohn}}{{Valluri} et~al.}{2007}]{valluri2007}
{Valluri} M.,  {Vass} I.~M.,  {Kazantzidis} S.,  {Kravtsov} A.~V.,   {Bohn}
  C.~L.,  2007, \mn@doi [\apj] {10.1086/511298}, \href
  {http://adsabs.harvard.edu/abs/2007ApJ...658..731V} {658, 731}

\bibitem[\protect\citeauthoryear{{Vass}, {Kazantzidis}, {Valluri}  \&
  {Kravtsov}}{{Vass} et~al.}{2009}]{vass2009}
{Vass} I.~M.,  {Kazantzidis} S.,  {Valluri} M.,   {Kravtsov} A.~V.,  2009,
  \mn@doi [\apj] {10.1088/0004-637X/698/2/1813}, \href
  {http://adsabs.harvard.edu/abs/2009ApJ...698.1813V} {698, 1813}

\bibitem[\protect\citeauthoryear{{Vera-Ciro}, {Helmi}, {Starkenburg}  \&
  {Breddels}}{{Vera-Ciro} et~al.}{2013}]{veraciro2013}
{Vera-Ciro} C.~A.,  {Helmi} A.,  {Starkenburg} E.,   {Breddels} M.~A.,  2013,
  \mn@doi [\mnras] {10.1093/mnras/sts148}, \href
  {http://adsabs.harvard.edu/abs/2013MNRAS.428.1696V} {428, 1696}

\bibitem[\protect\citeauthoryear{{Vera-Ciro}, {Sales}, {Helmi}  \&
  {Navarro}}{{Vera-Ciro} et~al.}{2014}]{veraciro2014}
{Vera-Ciro} C.~A.,  {Sales} L.~V.,  {Helmi} A.,   {Navarro} J.~F.,  2014,
  \mn@doi [\mnras] {10.1093/mnras/stu153}, \href
  {http://adsabs.harvard.edu/abs/2014MNRAS.439.2863V} {439, 2863}

\bibitem[\protect\citeauthoryear{{Wong} \& {Taylor}}{{Wong} \&
  {Taylor}}{2012}]{wong2012}
{Wong} A.~W.~C.,  {Taylor} J.~E.,  2012, \mn@doi [\apj]
  {10.1088/0004-637X/757/1/102}, \href
  {http://adsabs.harvard.edu/abs/2012ApJ...757..102W} {757, 102}

\bibitem[\protect\citeauthoryear{{Zemp}, {Moore}, {Stadel}, {Carollo}  \&
  {Madau}}{{Zemp} et~al.}{2008}]{zemp2008}
{Zemp} M.,  {Moore} B.,  {Stadel} J.,  {Carollo} C.~M.,   {Madau} P.,  2008,
  \mn@doi [\mnras] {10.1111/j.1365-2966.2008.13126.x}, \href
  {http://adsabs.harvard.edu/abs/2008MNRAS.386.1543Z} {386, 1543}

\bibitem[\protect\citeauthoryear{{Zhao}, {Mo}, {Jing}  \& {B{\"o}rner}}{{Zhao}
  et~al.}{2003}]{zhao2003a}
{Zhao} D.~H.,  {Mo} H.~J.,  {Jing} Y.~P.,   {B{\"o}rner} G.,  2003, \mn@doi
  [\mnras] {10.1046/j.1365-8711.2003.06135.x}, \href
  {http://adsabs.harvard.edu/abs/2003MNRAS.339...12Z} {339, 12}

\bibitem[\protect\citeauthoryear{{Zhao}, {Jing}, {Mo}  \& {B{\"o}rner}}{{Zhao}
  et~al.}{2009}]{zhao2009}
{Zhao} D.~H.,  {Jing} Y.~P.,  {Mo} H.~J.,   {B{\"o}rner} G.,  2009, \mn@doi
  [\apj] {10.1088/0004-637X/707/1/354}, \href
  {http://adsabs.harvard.edu/abs/2009ApJ...707..354Z} {707, 354}

\bibitem[\protect\citeauthoryear{{van den Bosch}}{{van den
  Bosch}}{2002}]{vandenbosch2002}
{van den Bosch} F.~C.,  2002, \mn@doi [\mnras]
  {10.1046/j.1365-8711.2002.05171.x}, \href
  {http://adsabs.harvard.edu/abs/2002MNRAS.331...98V} {331, 98}

\makeatother
\end{thebibliography}
% Don't change these lines
\bsp	% typesetting comment
\label{lastpage}
\end{document}